\documentclass[prc,notitlepage, twocolumn,
showpacs,floatfix,nofootinbib,preprintnumbers,superscriptaddress,amsmath,amssymb]{revtex4-1}

\usepackage{graphicx}\usepackage{epsfig}
\usepackage{amsmath,amssymb}
\usepackage{bm}
\usepackage{color}
\usepackage{float}
\usepackage{dcolumn} 
\usepackage{enumerate}
\usepackage{multirow}
\usepackage{threeparttable}
\usepackage{slantsc}

\begin{document}

\newcommand{\ket}[1]{\left| #1 \right\rangle}
\newcommand{\bra}[1]{\left\langle #1 \right|}
\newcommand{\Fhat}{\hat{F}}

\title{Collective inertia of Nambu-Goldstone mode from linear response theory}

\author{Nobuo Hinohara}
\email{hinohara@ccs.tsukuba.ac.jp}
\affiliation{%
Center for Computational Sciences, University of Tsukuba, Tsukuba, 305-8577, Japan
}
\affiliation{%
National Superconducting Cyclotron Laboratory, Michigan State University, East Lansing, Michigan, 48824-1321, USA
}

\begin{abstract}
  \begin{description}
  \item[Background]
    Spurious zero-energy Nambu-Goldstone (NG) modes appear when the symmetry of a system is spontaneously broken. The Thouless-Valatin inertia, the collective inertia of the NG mode, contains important information concerning collective motion.
  \item[Purpose]
    To establish an efficient and precise method for deriving the collective inertia and the conjugate operator of a NG mode, we derive an expression for the response function in terms of the coordinate-momentum representation of the quasiparticle random-phase approximation which is valid even if a symmetry-restoring zero-energy mode is present. 
\item[Methods]
  We use the finite amplitude method for computing the response function of superfluid nuclei with the nuclear density functional theory.
\item[Results]
  We derived analytically the collective inertia and the conjugate coordinate operator of the NG mode from the zero-energy linear response with the momentum operator of the NG mode.
  The formulation is tested in the cases of translational and pairing rotational modes. Illustrative calculations are performed for the neutron pairing rotation in Sn isotopes, the proton pairing rotation in $N=82$ isotones, and the neutron and proton pairing rotations around the $^{130}$Xe nucleus.
 \item[Conclusions]
   The proposed formulation allows us to compute the collective inertia of the NG mode precisely and efficiently.
   The conjugate coordinate operator can be utilized to remove spurious contributions to the strength distribution in the finite amplitude method.
  \end{description}
  \end{abstract}


\maketitle

\section{Introduction}

Spontaneous symmetry breaking is a universal phenomenon that plays an essential role  in various fields of physics. The emergence of pions from chiral symmetry breaking and gauge-symmetry breaking in superconducting systems are typical examples.

The nucleus is a finite quantum system, whose exact ground state does not break the symmetries of the Hamiltonian.
However, if we introduce a one-body mean-field approximation such as the Hartree-Fock-Bogoliubov (HFB) approximation to the low-energy nuclear many-body problem, spontaneous symmetry breaking can take place to account for more correlations within the one-body approximation \cite{Nazarewicz199427,RevModPhys.73.463, 0034-4885-68-1-R03, 0034-4885-70-12-R02,Birman20131}.
Continuous symmetries that are conserved in the nuclear Hamiltonian which can be broken spontaneously are the translational, rotational, and particle-number gauge symmetries. The isospin symmetry can be broken spontaneously \cite{1402-4896-2000-T88-032,PhysRevC.80.044313}, but is explicitly broken in the level of the effective nuclear and Coulomb interactions \cite{PhysRevLett.106.132502}.

If a continuous symmetry is broken spontaneously, a zero-energy Nambu-Goldstone (NG) mode appears which restores it \cite{PhysRev.117.648,INC_19_154}.
In the case of the symmetry-broken nuclear mean field, the NG mode appears in the self-consistent solution to the quasiparticle random-phase approximation (QRPA) as a symmetry restoration mode \cite{Ring-Schuck}.
The NG modes that correspond to broken symmetries in the mean field approximation
(for example: the center of mass mode for translational symmetry, the rotational mode for rotational symmetry, and the pairing rotational mode for particle-number gauge symmetry)
are associated with infinitesimal transformations of the frame of reference. As such, NG modes are known as ``spurious modes,'' since they do not represent a physical excitation within the intrinsic frame.

The actual spectroscopy measurement is performed in the laboratory frame for a nuclear energy eigenstate that preserves the symmetries of the Hamiltonian. Then the inertia of the NG mode in the intrinsic frame has an experimental correspondence.
The collective inertia for the NG mode within the QRPA framework is called the Thouless-Valatin inertia \cite{Thouless1962211}.
The meaning of this inertia changes depending on the NG mode present: for translational motion it represents the total mass of the nucleus; for the rotational and pairing-rotational modes, it represents the rotational and pairing-rotational moments of inertia, respectively.
Except for the center of mass mode where both the coordinate and momentum QRPA phonon operators are known \cite{Ring-Schuck}, the Thouless-Valatin inertia is not known in advance,
and the QRPA equations must be solved in order to find it.

The Thouless-Valatin inertia is important because it contains information on symmetry restoration, namely, ground state correlations to a particular broken symmetry \cite{Kammuri01061967, PAN64_588}.
This inertia was recently used for the three-dimensional spacial rotational mode in the five-dimensional quadrupole collective Hamiltonian~\cite{0954-3899-36-12-123101,PhysRevC.82.064313}.
When compared with the Belyaev moment of inertia \cite{PhysRev.103.1786,Belyaev196517} (its simplified version), the Thouless-Valatin inertia is typically 30\% higher; this is due to the contributions from two-body residual interactions not found in the Belyaev moment of inertia \cite{PhysRevC.60.054301}.
The Thouless-Valatin inertia has also been applied to symmetry breaking in quantum dots \cite{PhysRevB.68.035341}. Given its usefulness,
a microscopic derivation of the inertia based on the nuclear energy density functional (EDF) would be helpful in making systematic calculations with predictive power \cite{bertsch:032502}.

The expression for the Thouless-Valatin inertia is well known in terms of the QRPA $A$ and $B$ matrices \cite{Ring-Schuck, yoshida:044312,yoshida:064316,PhysRevC.83.061302}.
However, its direct evaluation is not trivial due to the size of these matrices.
Therefore, several alternative approaches for computing the Thouless-Valatin inertia have been proposed: the cranked mean-field calculation \cite{PhysRevC.60.054301,PhysRevC.62.054306,PhysRevC.81.014303} and a perturbative expansion based on adiabatic time-dependent HFB \cite{PhysRevC.86.034334}.

An alternative approach to solving the QRPA equations based on linear response theory has been developed for nuclear density functional theory: 
it is called the finite amplitude method (FAM) \cite{nakatsukasa:024318}.
In the FAM, the response function to an external one-body field can be evaluated only through a one-body induced field, and the calculation cost is significantly reduced compared with the matrix diagonalization of the QRPA equations. The FAM has been implemented on various versions of Skyrme-HFB codes
\cite{PhysRevC.84.041305,PhysRevC.84.014314,PhysRevC.90.051304} and within covariant density functional theory \cite{PhysRevC.87.054310,PhysRevC.88.044327}.
Not only does the FAM serve as an efficient way to calculate strength functions \cite{PhysRevC.84.021302,PhysRevC.89.064316}, it has also been used to improve upon solution methods of the QRPA equations \cite{PhysRevC.87.014331,PhysRevC.87.064309,PhysRevC.91.044323}.
Up to now the FAM has been used to find physical excitations, and has not been formulated for the symmetry restoration NG modes.

The NG mode sometimes gives an unphysical contribution to a physical excitation that has the same quantum numbers.
Its energy can be non-zero when a self consistency between the HFB and QRPA equations is broken, or when basis truncation violates the exact symmetry numerically.
In Ref.~\cite{nakatsukasa:024318}, a prescription to remove such contamination from the FAM response function is proposed and demonstrated in the case of center of mass motion.
A similar prescription is employed in the iterative Arnoldi diagonalization \cite{PhysRevC.81.034312}.
To apply this prescription to other modes, however, 
we need full solutions of the NG mode, which are not known in advance other than in the case of center of mass mode.

The aim of this paper is to present a formalism of the FAM that can be applicable in the presence of the NG mode and give expressions for the Thouless-Valatin inertia and conjugate coordinate operator of the NG modes.
The expression of the FAM response function in terms of the QRPA solutions \cite{PhysRevC.87.064309} is based on the $XY$ representation of the QRPA equation, and is not applicable to the NG mode.
In this paper, we derive the spectral representation of the response function of the FAM in terms of the momentum-coordinate ($PQ$) representation of the QRPA \cite{Marshalek1969,Ring-Schuck}. This is done to handle both the NG modes and imaginary solutions of the QRPA in addition to the physical modes.

This paper is organized as follows. In Sec.~\ref{sec:QRPA}, the QRPA equations in the $PQ$ representation are recapitulated. Section~\ref{sec:FAM} contains a brief introduction to the FAM, and its response function is expressed in the $PQ$ representation of the QRPA in Sec.~\ref{sec:FAMPQ}.
Then in Sec.~\ref{sec:mass}, the expression for the Thouless-Valatin inertia is derived. Numerical tests of the formalism for the center of mass motion and pairing rotations are presented in Sec.~\ref{sec:numerical}, and
realistic calculations of the neutron and proton pairing rotations are shown in Sec.~\ref{sec:pairrot}.
Lastly, conclusions and outlook are given in Sec.~\ref{sec:conclusion}.

\section{QRPA in $XY$ and $PQ$ representations \label{sec:QRPA}}
We recapitulate the QRPA matrix expression for later convenience \cite{Ring-Schuck}. The QRPA equation in $XY$ representation is given by
\begin{align}
  {\cal S} {\cal X} = \Sigma_3{\cal X} {\cal O},
\end{align}
where
\begin{align}
  {\cal S}_{\mu\nu,\mu'\nu'} =& \begin{pmatrix} A & B \\ B^\ast & A^\ast \end{pmatrix}_{\mu\nu,\mu'\nu'},  &
  {\cal X}_{\mu\nu,i} =& 
  \begin{pmatrix} X^i & Y^{i\ast} \\ Y^i & X^{i\ast} \end{pmatrix}_{\mu\nu},\\
  \Sigma_3 =&
  \begin{pmatrix} 1 & 0 \\ 0 & -1 \end{pmatrix}, \quad &
  {\cal O}_{ij} =&
  \begin{pmatrix} \Omega & 0 \\ 0 & -\Omega \end{pmatrix}_{ij},
\end{align}
where $\mu\nu$, $\mu'\nu'$ are the two-quasiparticle indices, $ij$ are the indices for QRPA eigenmodes,
$A$ and $B$ are the QRPA matrices, and $\Omega_i$ is the energy of the QRPA eigenmode.
The indices of $\Sigma_3$ can be either $\mu\nu,\mu'\nu'$ or $ij$.
The matrix $\Omega$ in ${\cal O}$ is $\Omega_{ij} = \Omega_i \delta_{ij}$.
The amplitudes $X^i$ and $Y^i$ are the two-quasiparticle amplitudes of a QRPA phonon operator
\begin{align}
  \hat{O}^\dagger_i = \sum_{\mu<\nu} X^i_{\mu\nu} \hat{\bm{A}}^\dag_{\mu\nu}- Y^i_{\mu\nu} \hat{\bm{A}}_{\mu\nu},
\end{align}
with $\hat{\bm{A}}^\dag_{\mu\nu}=  \hat{a}^\dag_\mu \hat{a}^\dag_\nu$;
they are normalized with
\begin{align}
  \langle [\hat{O}_i,\hat{O}_j^\dag]\rangle =& \delta_{ij},  \quad 
    \langle [\hat{O}^\dag_i,\hat{O}^\dag_j]\rangle = 0,  \quad 
\end{align}
or equivalently
\begin{align}
  {\cal X}^\dag \Sigma_3{\cal X} = \Sigma_3, \quad
  {\cal X}\Sigma_3{\cal X}^\dagger = \Sigma_3.
  \label{eq:XYnorm}
\end{align}
The QRPA equations are written in the Hamiltonian expression as
\begin{align}
  [ \hat{H}_{\rm QRPA}, \hat{O}_i^\dag] = \Omega_i \hat{O}_i^\dag, \quad
  [ \hat{H}_{\rm QRPA}, \hat{O}_i] = -\Omega_i \hat{O}_i,
\end{align}
where $\hat{H}_{\rm QRPA}$ is the QRPA part of the Hamiltonian.
We always have pairs of solutions with positive and negative frequencies.
This expression assures that the two-quasiparticle part of the Hermitian broken-symmetry operator $\hat{P}_{\rm NG}$ given by
\begin{align}
  \hat{P}_{\rm NG}= \sum_{\mu<\nu} (P_{\rm NG})_{\mu\nu} \hat{\bm{A}}^\dag_{\mu\nu}
  + (P_{\rm NG})^\ast_{\mu\nu} \hat{\bm{A}}_{\mu\nu}
\end{align}
is always a solution $\hat{O}_{\rm NG}$ of the equation with $\Omega_{\rm NG}=0$.
The problem with the $XY$ representation of the QRPA equations for NG modes is that  
they cannot be normalized with Eq.~(\ref{eq:XYnorm}).

We now switch to the $PQ$ representation to handle NG modes.
We introduce Hermitian coordinate and momentum operators that describe the eigenmodes as
\begin{align}
  \hat{\cal Q}_i =& \sqrt{ \frac{1}{2M_i\Omega_i}} ( \hat{O}_i + \hat{O}_i^\dag )
  = \sum_{\mu<\nu} Q^i_{\mu\nu} \hat{\bm{A}}^\dag_{\mu\nu} + Q^{i\ast}_{\mu\nu} \hat{\bm{A}}_{\mu\nu}, \\
  \hat{\cal P}_i =& \frac{1}{i} \sqrt{ \frac{ M_i\Omega_i}{2}} ( \hat{O}_i - \hat{O}_i^\dag )
  = \sum_{\mu<\nu} P^i_{\mu\nu} \hat{\bm{A}}^\dag_{\mu\nu} + P^{i\ast}_{\mu\nu} \hat{\bm{A}}_{\mu\nu},
\end{align}
and regard these operators as a starting point. Here $M_i$ is the inertia for each mode.
The QRPA equations in terms of these operators are written as
\begin{align}
  \sum_{\mu'<\nu'}
  \begin{pmatrix} A & B \\ B^\ast & A^\ast \end{pmatrix}_{\mu\nu,\mu'\nu'}
    \begin{pmatrix} P_i \\ -P_i^{\ast} \end{pmatrix}_{\mu'\nu'}
    =&
    i \Omega_i^2 M_i
    \begin{pmatrix} Q_i \\ Q_i^\ast \end{pmatrix}_{\mu\nu},
    \\ 
    \sum_{\mu'<\nu'}
      \begin{pmatrix} A & B \\ B^\ast & A^\ast \end{pmatrix}_{\mu\nu,\mu'\nu'}
    \begin{pmatrix} Q_i \\ -Q_i^{\ast} \end{pmatrix}_{\mu'\nu'}
    =&
    -\frac{i}{M_i}
    \begin{pmatrix} P_i \\ P_i^\ast \end{pmatrix}_{\mu\nu}.
    \end{align}
By defining the matrices
\begin{align}
  {\cal V}_{\mu\nu,i} = \begin{pmatrix} P^i & Q^i \\ -P^{i\ast} & -Q^{i\ast} \end{pmatrix}_{\mu\nu,i},
  {\cal W}_{ij} = \begin{pmatrix} 0 & -i M^{-1} \\ i M\Omega^2 & 0 \end{pmatrix}_{ij}, 
\end{align}
with $M_{ij}=M_i\delta_{ij}$,
the QRPA equations in $PQ$ representation are summarized as
\begin{align}
  {\cal S}{\cal V} = \Sigma_3{\cal V}{\cal W}. \label{eq:QRPAinPQ}
  \end{align}
The operators $\hat{\cal P}_i$ and $\hat{\cal Q}_j$ are normalized with
\begin{align}
  \langle [ \hat{\cal Q}_i, \hat{\cal P}_j]\rangle =& i\delta_{ij}, &\quad
    \langle [ \hat{\cal Q}_i, \hat{\cal Q}_j]\rangle =  
  \langle [ \hat{\cal P}_i, \hat{\cal P}_j]\rangle =& 0, \label{eq:PQnorm}
  \end{align}
or equivalently
\begin{align}
  {\cal V}^\dag\Sigma_3{\cal V} = \Sigma_2, \quad
  {\cal V}\Sigma_2{\cal V}^\dagger = \Sigma_3, \label{eq:PQnorm2}
\end{align}
where
\begin{align}
  \Sigma_2 = \begin{pmatrix} 0 & -i \\ i & 0 \end{pmatrix}.
\end{align} 
The solution for a zero-energy NG mode is written as
\begin{align}
  \begin{pmatrix} A & B \\ B^\ast & A^\ast \end{pmatrix}
  \begin{pmatrix} P_{\rm NG} \\ -P^\ast_{\rm NG} \end{pmatrix}
  =& 0, \\
  \begin{pmatrix} A & B \\ B^\ast & A^\ast \end{pmatrix}
  \begin{pmatrix} Q_{\rm NG} \\ -Q^\ast_{\rm NG} \end{pmatrix}
  =& -\frac{i}{M_{\rm NG}}
  \begin{pmatrix} P_{\rm NG} \\ P^\ast_{\rm NG} \end{pmatrix},
\end{align}
where $M_{\rm NG}$ is the Thouless-Valatin inertia for the NG mode.
In the $PQ$ representation, $P_{\rm NG}$ and $Q_{\rm NG}$ are normalizable with Eq.~(\ref{eq:PQnorm})
and the Thouless-Valatin inertia is given by
\begin{align}
  M_{\rm NG} =
\begin{cases}
  2 P_{\rm NG}(A+B)^{-1}P_{\rm NG} \quad({\rm Im}\, P_{\rm NG}=0),\\
  -2 P_{\rm NG}(A-B)^{-1}P_{\rm NG} \quad({\rm Re}\, P_{\rm NG}=0).
\end{cases} \label{eq:TVMOI-QRPA}
\end{align}
The expression above depends on whether $P_{\rm NG}$ is real or pure imaginary.
Computation of the Thouless-Valatin inertia from Eq.~(\ref{eq:TVMOI-QRPA})
requires full evaluation of the $A$ and $B$ matrices of large dimensions
for recent nuclear density functional theory without symmetry restrictions.
In the next section, we derive expressions for the Thouless-Valatin inertia based on linear response theory.

\section{Finite-amplitude method \label{sec:FAM}}
In this section we introduce the FAM, and express the response function in the $PQ$ representation following the notations in Ref.~\cite{PhysRevC.84.014314}.
We start with an external time-dependent field $\hat{F}(t)$ with a frequency $\omega$ and a small finite amplitude parameter $\eta$
\begin{align}
  \hat{F}(t) = \eta \left\{ \hat{F} e^{-i\omega t} + \hat{F}^\dag e^{i\omega t}\right\}
\end{align}
applied to the system. Here the one-body operator is written in the quasiparticle basis
\begin{align}
  \hat{F} = \sum_{\mu<\nu} \left(F^{20}_{\mu\nu} \hat{\bm{A}}^\dag_{\mu\nu} + F^{02}_{\mu\nu}\hat{\bm{A}}_{\mu\nu}\right) + \sum_{\mu\nu} F^{11}_{\mu\nu}\hat{\bm{B}}_{\mu\nu},
\end{align}
where $\hat{\bm{B}}_{\mu\nu}=\hat{a}^\dag_\mu \hat{a}_\nu$.
In the FAM, we solve the time-dependent HFB (TDHFB) equations with the external field
\begin{align}
  i \frac{\partial}{\partial t} \hat{a}_{\mu}(t) = [ \hat{H}(t) + \hat{F}(t),
    \hat{a}_\mu(t)].
\end{align}
The time dependence is governed by the forced oscillation of the external field.
The time-dependence of the quasiparticle and the Hamiltonian is given by
\begin{align}
  \hat{a}_\mu(t) =& \left\{\hat{a}_\mu + \delta \hat{a}_\mu(t) \right\}e^{iE_\mu t},\\
  \delta \hat{a}_\mu(t) =& \eta \sum_{\nu}
  \hat{a}^\dag_{\nu} \left\{ X_{\nu\mu}(\omega) e^{-i\omega t} + Y^\ast_{\nu\mu}(\omega) e^{i\omega t}\right\}, \\
  \hat{H}(t) =& \sum_{\mu} E_\mu \hat{\bm{B}}_{\mu\mu} + \delta \hat{H}(t), \\
  \delta \hat{H}(t) =& \eta \left\{ \delta \hat{H}(\omega) e^{-i\omega t}
  + \delta \hat{H}^\dag(\omega) e^{i\omega t}\right\}, \\
  \delta \hat{H}(\omega)=& \sum_{\mu<\nu}\left\{
  \delta H^{20}_{\mu\nu}(\omega) \hat{\bm{A}}^\dag_{\mu\nu} +
  \delta H^{02}_{\mu\nu}(\omega) \hat{\bm{A}}_{\mu\nu}
  \right\},
\end{align}
where $E_\mu$ is the quasiparticle energy and $X_{\mu\nu}(\omega)$ and $Y_{\mu\nu}(\omega)$ are the FAM amplitudes.
Using the expressions above and taking the terms linear to $\eta$ (small-amplitude approximation), the TDHFB equations are written as
\begin{align}
  (E_\mu + E_\nu - \omega) X_{\mu\nu}(\omega) + \delta H^{20}_{\mu\nu}(\omega) =& - F^{20}_{\mu\nu}, \\
  (E_\mu + E_\nu + \omega) Y_{\mu\nu}(\omega) + \delta H^{02}_{\mu\nu}(\omega) =& - F^{02}_{\mu\nu}.
\end{align}
This expression does not involve the $A$ and $B$ matrices explicitly, but through the one-body induced fields, $\delta H^{20}(\omega)$ and $\delta H^{02}(\omega)$. This is the advantage of the FAM: the response of the system to the external field $\hat{F}$ can be evaluated from one-body quantities only. We refer more details on how to compute the one-body induced field in nuclear density functional theory to Ref.~\cite{PhysRevC.84.014314}. We also note that the FAM equation is formally equivalent to linear response theory, and can be written as
\begin{align}
  \begin{pmatrix} X(\omega) \\ Y(\omega) \end{pmatrix} =& -
  \left[ \begin{pmatrix} A & B \\ B^\ast  & A^\ast \end{pmatrix}
    - \omega \begin{pmatrix} 1 & 0 \\ 0 & -1 \end{pmatrix}
    \right]^{-1}
  \begin{pmatrix} F^{20} \\ F^{02}\end{pmatrix} \nonumber \\
  =& -R(\omega)   \begin{pmatrix} F^{20} \\ F^{02}\end{pmatrix}, \label{eq:FAMeqAB}
\end{align}
where $R(\omega)$ is the response function.

\section{FAM response function in the $PQ$ representation \label{sec:FAMPQ}}

Now we derive the expression of the response function $R(\omega)$ in terms of the $PQ$ representation of the QRPA. A similar derivation has been done for the $XY$ representation of the QRPA in Ref.~\cite{PhysRevC.87.064309}.
Using Eqs.~(\ref{eq:QRPAinPQ}) and (\ref{eq:PQnorm2}), the response function $R(\omega)$ is written as
\begin{align}
  R(\omega) =& [ {\cal S} - \omega \Sigma_3]^{-1}
  = {\cal V} [ {\cal W} - \omega{\cal I}]^{-1} \Sigma_2 {\cal V}^\dag,
\end{align}
where
\begin{align}
  [{\cal W} - \omega {\cal I}]^{-1} =
  \begin{pmatrix} -\omega & - iM^{-1} \\ i M \Omega^2 & -\omega \end{pmatrix}^{-1},
\end{align}
and ${\cal I}$ is a unit matrix.
We note that this expression cannot be defined at $\omega=\pm \Omega_i$.
This excludes $\omega=0$ if a NG mode is present.
This matrix has blocked structure for the same index in the four blocks, and we
can take the inverse for each $2\times 2$ matrix.
The response function and the FAM amplitudes with $P$ and $Q$ coefficients are now given as
\begin{widetext}
\begin{align}
 &R(\omega)_{\mu\nu,\mu\nu'} = \sum_{i} \frac{1}{\omega^2 - \Omega_i^2} \nonumber \\
 &\begin{bmatrix} 
    (i\omega P^i_{\mu\nu} - M_i\Omega_i^2 Q_{\mu\nu}^i )Q^{i\ast}_{\mu'\nu'}
    +(- M_i^{-1} P^i_{\mu\nu} - i\omega Q_{\mu\nu}^i )P^{i\ast}_{\mu'\nu'} &
    (-i\omega P^i_{\mu\nu}+ M_i\Omega_i^2 Q_{\mu\nu}^i) Q_{\mu'\nu'}^{i}
    + (M_i^{-1} P^i_{\mu\nu} +i\omega Q_{\mu\nu}^i) P^{i}_{\mu'\nu'} \\
    (-i\omega P^{i\ast}_{\mu\nu} + M_i\Omega_i^2 Q_{\mu\nu}^{i\ast}) Q_{\mu'\nu'}^{i\ast}
    + (M_i^{-1} P^{i\ast}_{\mu\nu} + i\omega Q_{\mu\nu}^{i\ast}) P^{i\ast}_{\mu'\nu'} & 
    (i\omega P^{i\ast}_{\mu\nu} - M_i\Omega_i^2 Q_{\mu\nu}^{i\ast}) Q_{\mu'\nu'}^{i}
    +(- M_i^{-1} P^{i\ast}_{\mu\nu} - i\omega Q_{\mu\nu}^{i\ast}) P^{i}_{\mu'\nu'} 
   \end{bmatrix}, \label{eq:responsefunc}\\
&  \begin{bmatrix} X_{\mu\nu}(\omega) \\ Y_{\mu\nu}(\omega) \end{bmatrix}
  = -\sum_{\mu'\nu} R_{\mu\nu\mu'\nu'}(\omega)
  \begin{bmatrix} F^{20}_{\mu'\nu'} \\ F^{02}_{\mu'\nu'} \end{bmatrix}
  = \sum_i\frac{1}{\omega^2 - \Omega_i^2}
   \begin{bmatrix}
    \displaystyle(-i\omega P^i_{\mu\nu}+M_i\Omega_i^2 Q^i_{\mu\nu}) \langle Q_i|\hat{F}|0\rangle + 
    \left(\frac{1}{M_i} P^i_{\mu\nu} + i\omega Q^i_{\mu\nu}\right) \langle P_i|\hat{F}|0\rangle \\ 
    \displaystyle(i\omega P^{i\ast}_{\mu\nu}-M_i\Omega_i^2 Q^{i\ast}_{\mu\nu}) \langle Q_i|\hat{F}|0\rangle
    + \left(-\frac{1}{M_i} P^{i\ast}_{\mu\nu} -i\omega Q^{i\ast}_{\mu\nu}\right) \langle P_i|\hat{F}|0\rangle
    \end{bmatrix},
\end{align}
\end{widetext}
where we define the transition strengths
from the ground state to the states expressed with $\hat{\cal P}_i$ and $\hat{\cal Q}_i$ phonon operators as
\begin{align}
  \langle P_i|\hat{F}|0\rangle \equiv & \langle [\hat{\cal P}_i, \hat{F}]\rangle
  = \sum_{\mu<\nu}  P^{i\ast}_{\mu\nu} F^{20}_{\mu\nu} - P^i_{\mu\nu} F^{02}_{\mu\nu}, \\
    \langle Q_i|\hat{F}|0\rangle \equiv &\langle [\hat{\cal Q}_i, \hat{F}]\rangle
  = \sum_{\mu<\nu}  Q^{i\ast}_{\mu\nu} F^{20}_{\mu\nu} - Q^i_{\mu\nu} F^{02}_{\mu\nu}. 
\end{align}
We note that the FAM amplitudes $X(\omega)$ and $Y(\omega)$ are not $X^i$ and $Y^i$ eigenvectors of the QRPA themselves, therefore they are well-defined through the linear response equation even if there are NG modes.
The FAM strength function is
\begin{align}
  S(\hat{F},\omega) =& \sum_{\mu<\nu} F^{20\ast}_{\mu\nu}X_{\mu\nu}(\omega) + F^{02\ast}_{\mu\nu} Y_{\mu\nu}(\omega) \nonumber \\
  =& \sum_i \frac{1}{\omega^2 - \Omega_i^2}\left\{
  \frac{1}{M_i} |\langle P_i|\hat{F}|0\rangle|^2
  +M_i\Omega_i^2 |\langle Q_i|\hat{F}|0\rangle|^2 \right.\nonumber  \\
& \left.  +\omega [QP]_i(\hat{F})\right\}, \label{eq:SFwinPQ}
\end{align}
where we define a real quantity $[QP]_i(\hat{F})\equiv i\left( \langle Q_i|\hat{F}|0\rangle^\ast\langle P_i|\hat{F}|0\rangle
-\langle P_i|\hat{F}|0\rangle^\ast\langle Q_i|\hat{F}|0\rangle \right)$.
When NG modes are not present ($\Omega_i\ne 0$), the following transition strength can be defined:
\begin{align}
  \langle i| \hat{F}|0\rangle =& i \sqrt{\frac{1}{2 M_i\Omega_i}} \langle P_i|\hat{F}|0\rangle
  +  \sqrt{ \frac{M_i\Omega_i}{2}} \langle Q_i|\hat{F}|0\rangle,
    \\ 
  \langle 0| \hat{F}|i\rangle =& i \sqrt{\frac{1}{2 M_i\Omega_i}} \langle P_i|\hat{F}|0\rangle
  -  \sqrt{ \frac{M_i\Omega_i}{2}} \langle Q_i|\hat{F}|0\rangle,
\end{align}
and by substituting these into Eq.~(\ref{eq:SFwinPQ}), we can go back to the original expression of the FAM strength function \cite{PhysRevC.87.064309}
\begin{align}
  S(\hat{F},\omega) = - \sum_{i>0} \left(
  \frac{|\langle i|\hat{F}|0\rangle|^2}{\Omega_i - \omega}
  +   \frac{|\langle 0|\hat{F}|i\rangle|^2}{\Omega_i + \omega}
  \right).
\end{align}
Generally, the solutions of the QRPA equations consist of physical modes with $\Omega_i^2>0$, NG modes with $\Omega_i^2=0$, and imaginary modes with $\Omega_i^2<0$. Imaginary solutions of the QRPA equations can occur when the HFB state does not correspond to a variational minimum. The $PQ$ representation of the strength function (\ref{eq:SFwinPQ}) is valid for all three kinds of the modes.
We consider the QRPA equations at a non-variational minimum mainly in two cases.
When the HFB code has a symmetry restriction, we cannot take the variation against the restricted degrees of freedom, and unexpectedly the HFB state obtained with the symmetry-restricted code can be unstable. Typical examples are when a deformed nucleus is computed with a spherical HFB code, or when a triaxial state is computed with an axial HFB code. The transition to isoscalar pairing condenstation with a HFB code with proton-neutron symmetry and proton-neutron particle-particle RPA was recently discussed from this point of view \cite{PhysRevC.90.031303}.
Another case is large-amplitude collective motion. The local QRPA \cite{PhysRevC.82.064313} based on the adiabatic theory of large-amplitude collective motion \cite{PTP.103.959} requires the solutions of the QRPA equations at non-equilibrium HFB states.

We can split the contributions of three modes to the strength function as
\begin{align}
  S(\hat{F},\omega) =& S(\hat{F},\omega)_{\rm phys} + S(\hat{F},\omega)_{\rm NG} + S(\hat{F},\omega)_{\rm imag},
\end{align}
where the contribution from the NG mode is given by
\begin{align}
  S(\hat{F},\omega)_{\rm NG} =&
  \sum_{i,\Omega_i=0} \left\{ 
  \frac{|\langle P_i|\hat{F}|0\rangle|^2}{M_i \omega^2} 
  +
  \frac{[QP]_i(\hat{F})}{\omega}
  \right\}. \label{eq:strengthNG}
\end{align}
The same expression is found in Ref.~\cite{Blaizot-Ripka}.
When we compute the strength function distribution with the FAM, we replace the real frequency $\omega$ by a complex value for the frequency $\omega+i\gamma$ where the imaginary part gives the width $\Gamma=2\gamma$.
If the external field can excite the NG mode, there is a spurious contribution
\begin{align}
  - \frac{1}{\pi} {\rm Im} S(F,&\omega+i\gamma)_{\rm NG} \nonumber \\
  =& \frac{\gamma}{\pi} \sum_{i,\Omega_i=0} \left\{
 \frac{2\omega |\langle P_i |\hat{F}|0\rangle|^2}{M_i(\omega^2+\gamma^2)^2}
 +
  \frac{[QP]_i(\hat{F})}{\omega^2+\gamma^2}
  \right\}
\end{align}
to the strength distribution.
The procedure to remove this contribution from the center of mass mode has been proposed in Ref.~\cite{nakatsukasa:024318}.

\section{Thouless-Valatin inertia for Nambu-Goldstone modes \label{sec:mass}}

\subsection{Thouless-Valatin inertia from the momentum operator}

The momentum operator $\hat{\cal P}_{\rm NG}$ of a NG mode is a consequence of a broken symmetry of the system. By using it as an external field of the FAM, we can show that the contribution to the strength function is zero
\begin{align}
  S(\hat{\cal P}_{\rm NG}, \omega) = 0
\end{align}
from Eqs.~(\ref{eq:PQnorm}), (\ref{eq:SFwinPQ}), and (\ref{eq:strengthNG}). Here we recall that the response function (\ref{eq:responsefunc}) is undefined at $\omega=0$ in the presence of the NG mode.
From Eq.~(\ref{eq:FAMeqAB}), the linear response equation at $\omega=0$ is written as
\begin{align}
  \begin{pmatrix} A & B \\ B^\ast & A^\ast \end{pmatrix}
  \begin{pmatrix} X(0) \\ Y(0) \end{pmatrix}
  =
  -
  \begin{pmatrix} P_{\rm NG} \\ P^\ast_{\rm NG} \end{pmatrix}.
\end{align}
Assuming that the $A$ and $B$ matrices are real, we have
\begin{align}
  (A+B)[X(0)+Y(0)] =& -2P_{\rm NG} \quad ({\rm Im}\, P_{\rm NG}=0),\\
  (A-B)[X(0)-Y(0)] =& -2P_{\rm NG} \quad ({\rm Re}\, P_{\rm NG}=0).
\end{align}
The FAM strength function at $\omega=0$ is then
\begin{align}
  S(\hat{\cal P}_{\rm NG}, \omega=0) =& \sum_{\mu<\nu}
  (P^\ast_{\rm NG})_{\mu\nu} X_{\mu\nu}(0) + (P^\ast_{\rm NG})_{\mu\nu} Y_{\mu\nu}(0)\nonumber \\ 
  =&\begin{cases}\displaystyle
    -2 P_{\rm NG} (A+B)^{-1} P_{\rm NG} \quad ({\rm Im}\, P_{\rm NG}=0)\\ \displaystyle
     2 P_{\rm NG} (A-B)^{-1} P_{\rm NG} \quad ({\rm Re}\, P_{\rm NG}=0)
  \end{cases} \nonumber \\
  =& - M_{\rm NG}, \label{eq:TVfromP}
\end{align}
where we use Eq.~(\ref{eq:TVMOI-QRPA}).
Therefore the FAM strength function for the momentum operator of the NG mode is
summarized as
\begin{align}
  S( \hat{\cal P}_{\rm NG}, \omega) = \begin{cases}
    0 \quad  &(\omega \ne \pm\Omega_i) \\
    - M_{\rm NG} \quad &(\omega=0) \label{eq:S_NG_P}
    \end{cases},
\end{align}
and the strength at zero frequency gives the Thouless-Valatin inertia.
The coordinate opeartor of the NG mode is then given by
\begin{align}
  Q_{\rm NG} =&
  \begin{cases}
    iM_{\rm NG}^{-1}(A+B)^{-1} P_{\rm NG}  \quad ({\rm Im}\, P_{\rm NG}=0)\\
   -i M_{\rm NG}^{-1}(A-B)^{-1} P_{\rm NG}  \quad ({\rm Re}\, P_{\rm NG}=0)\\
  \end{cases} \nonumber \\
 =& \begin{cases} 
   \displaystyle i\frac{X(0)+Y(0)}{2S(\hat{\cal P}_{\rm NG},0)}\quad ({\rm Im}\, P_{\rm NG}=0), \\
   \displaystyle i\frac{X(0)-Y(0)}{2S(\hat{\cal P}_{\rm NG},0)}\quad ({\rm Re}\, P_{\rm NG}=0).
   \end{cases} \label{eq:Q_NG}
\end{align}
Except for the trivial case of the center of mass mode, the coordinate operator of the NG mode is not known in advance, and this expression will be useful for removing spurious modes \cite{nakatsukasa:024318}.

\subsection{Thouless-Valatin inertia from the coordinate operator}

An alternative derivation of the Thouless-Valatin inertia is found
from the FAM calculation using the conjugate coordinate operator of the NG mode $\hat{\cal Q}_{\rm NG}$ as an external field. In this case,
the strength function is derived from Eqs.~(\ref{eq:PQnorm}) and (\ref{eq:strengthNG}) as
\begin{align}
  S(\hat{\cal Q}_{\rm NG},\omega) = \frac{1}{M_{\rm NG}\omega^2} \quad
  (\omega \ne \pm \Omega_i). \label{eq:SFwQ}
\end{align}
Therefore the Thouless-Valatin inertia is given from the energy-weighted sum rule \cite{Ring-Schuck} as 
\begin{align}
  M_{\rm NG}^{-1} = 2m_1(\hat{\cal Q}_{\rm NG}) = \frac{2}{2\pi i} \int_{A_1}
  \omega S(\hat{\cal Q}_{\rm NG},\omega) d\omega, \label{eq:TVfromQ}
\end{align}
where $A_1$ is the counterclockwise half circle in the complex energy plane from $\omega=-i R_{A_1}$ to $i R_{A_1}$ centered at the origin \cite{PhysRevC.91.044323}.

\subsection{Thouless-Valatin inertia and the $m_{-1}$ sum rule}

The expression for the Thouless-Valatin inertia (\ref{eq:TVMOI-QRPA}) is same as the inverse-energy weighted sum rule given in Ref.~\cite{PhysRevC.79.054329} but for the momentum operator of the NG mode.
Moreover the dielectric theorem \cite{Ring-Schuck, PhysRevC.79.054329} connects the inverse-energy weighted sum rule through the constrained HFB state $\ket{\phi(\lambda)}$ computed with the Hamiltonian $\hat{H} + \lambda \hat{\cal P}_{\rm NG}$
\begin{align}
  m_{-1}(\hat{\cal P}_{\rm NG}) = 
  - \frac{1}{2} \left[ \frac{\partial}{\partial \lambda}
    \langle \phi(\lambda)|\hat{\cal P}_{\rm NG}|\phi(\lambda)\rangle
  \right]_{\lambda=0} = -\frac{M_{\rm NG}}{2}.
\end{align} 
This theorem allows us to compute the Thouless-Valatin rotational moment of inertia from the cranked HFB calculation \cite{PhysRevC.62.054306, PhysRevC.65.041307, PhysRevC.60.054301,PhysRevB.68.035341}.
These facts show that the Thouless-Valatin inertia is related to the inverse-energy-weighted sum rule $m_{-1}(\hat{\cal P}_{\rm NG})$. However we need a careful consideration of the $m_{-1}$ sum rule when NG modes are present.
The expression of the $m_{-1}$ sum rule in terms of the transition matrix elements of an operator $\hat{F}$ is given by
\begin{align}
  m_{-1}(\hat{F}) = \sum_{i} \Omega_i^{-1} |\langle i|\hat{F}|0\rangle|^2,
\end{align}
and the NG modes should not be included in the summation.
Because of the discontinuous character of the FAM strength function $S(\hat{\cal P}_{\rm NG}, \omega=0)$, the inverse energy-weighted sum rule from the contour integration of Ref.~\cite{PhysRevC.91.044323} does not provide the Thouless-Valatin inertia.
The contribution from the NG mode to the $m_{-1}$ sum rule has to be added separately.
A similar discussion is found for the $m_1$ sum rule in the Appendix of Ref.~\cite{PhysRevC.67.044315}.

\subsection{Approximate symmetry \label{sec:approx}}
Even though the Hamiltonian preserves its symmetry, it can be explicitly broken in the settings of a numerical calculation.
This is actually the case for translational and rotational symmetries.
We often express the single-particle states either in the harmonic oscillator basis expanded about the center of mass or in the coordinate lattice of a finite box.
Ideally the translational displacement leaves the energy of the system invariant.
But if the wave function is expanded in a finite basis, the translational/rotational symmetry is broken explicitly.
This results in the translational and rotational modes appearing at finite energies.
We discuss the effect on such approximate symmetries on the evaluation of the Thouless-Valatin inertia. When the symmetry is approximate, the excitation energy $\Omega_{\rm NG}$ is finite, and the contribution to the strength function is actually the same as other physical modes. The momentum and coordinate operators of the NG mode
are now only approximate solutions to the QRPA equations.
Therefore we have the following approximate expressions for the strength function at energy around $\omega=\Omega_{\rm NG}$:
\begin{align}
 S(\hat{\cal Q}_{\rm NG},\omega) \sim & \frac{1}{M_{\rm NG}(\omega^2 - \Omega_{\rm NG}^2)},   \label{eq:SQapp} \\
 S(\hat{\cal P}_{\rm NG},\omega) \sim & \frac{M_{\rm NG} \Omega_{\rm NG}^2}{\omega^2-\Omega_{\rm NG}^2}. \label{eq:SPapp}
\end{align}
At $\omega=0$, the contribution from the finite excitation energy $\Omega_{\rm NG}$ in Eq.~(\ref{eq:SPapp}) is canceled, and Eq.~(\ref{eq:TVfromP}) remains as a good approximation to the Thouless-Valatin inertia.
The energy-weighted sum rule (\ref{eq:TVfromQ}) is valid if $R_{A_1}>\Omega_{\rm NG}$.
The position of the excitation energy of the NG mode is estimated from Eqs.~(\ref{eq:SQapp}) and (\ref{eq:SPapp}) as
\begin{align}
  \Omega_{\rm NG}^2 = \frac{1}{S(\hat{\cal P}_{\rm NG},0)S(\hat{\cal Q}_{\rm NG},0)}.
  \label{eq:NGenergy}
\end{align}

\subsection{Inglis-Belyaev cranking inertia}

The Inglis-Belyaev cranking inertia \cite{PhysRev.103.1786,Belyaev196517} is easily evaluated when the FAM routine is available.
The general expression of the cranking inertia for the Hermitian operators $\hat{F}_i$ and $\hat{F}_j$ is given by
\begin{align}
  M_{\rm IB}(ij) = 2\sum_{\mu<\nu} \frac{
    F^{i20\ast}_{\mu\nu} F^{j20}_{\mu\nu} + F^{i20}_{\mu\nu} F^{j20\ast}_{\mu\nu}}
    {E_\mu + E_\nu}. \label{eq:cranking}
\end{align}
By setting the induced field and the energy to zero in the FAM calculation
($\delta H^{20}_{\mu\nu} = \delta H^{02}_{\mu\nu} = 0$ and $\omega=0$),
the FAM amplitudes for operator $\hat{F}_i$ is given without self-consistent iteration:
\begin{align}
  X_{\mu\nu}(\hat{F}_i,\omega=0) = - \frac{F^{i,20}_{\mu\nu}}{E_\mu + E_\nu}, \\
  Y_{\mu\nu}(\hat{F}_i,\omega=0) = - \frac{F^{i,02}_{\mu\nu}}{E_\mu + E_\nu}.
\end{align}
The Inglis-Belyaev inertia is given from the amplitudes for another operator $\hat{F}_j$ as
\begin{align}
  \sum_{\mu<\nu} [F^{j, 20\ast}_{\mu\nu} X_{\mu\nu}(\hat{F}_i,0) +
  F^{j,02\ast}_{\mu\nu} Y_{\mu\nu}(\hat{F}_i,0)] = - M_{\rm IB}(ij).
\end{align}
If $\hat{F}_i=\hat{F}_j$, the Inglis-Belyaev inertia gives the response function
without an induced field.
This derivation agrees with the fact that the cranking inertia does not include the contribution from the residual interaction.

\section{Numerical tests \label{sec:numerical}}

To check the derivation in the previous section and to show its applicability, we performed numerical calculations for $^{26}$Mg with the EDF SLy4 \cite{Chabanat1998231} with volume pairing $V_0=-125.2$ MeV fm$^3$ and a 60 MeV quasiparticle energy cutoff.
The HFB and FAM calculations were based on the computer code {\sc HFBTHO} \cite{Stoitsov200543,Stoitsov20131592} and
its FAM extension to the $K=0$ mode \cite{PhysRevC.84.041305}.
The calculations were performed with various sizes of the harmonic oscillator space from $N_{\rm sh}=5$ to 20.
We used the proton-superconducting oblate deformed HFB state with $\beta=-0.18$ and $\Delta_p=0.619$ MeV. The broken symmetries in this HFB state are translational, rotational, and proton particle number.

\subsection{Center of mass mode}
\begin{figure}
  \includegraphics[width=80mm]{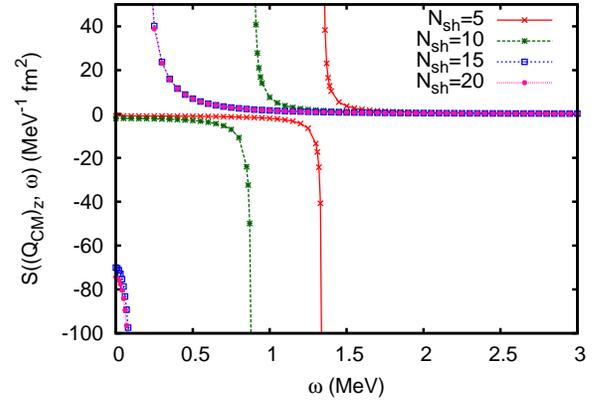}
  \caption{(Color online) The FAM strength function $S((\bm{\hat{Q}}_{\rm CM})_z,\omega)$ for a response to the center of mass coordinate operator with a real frequency $\omega$.}
  \label{fig:CM-Q}
\end{figure}

\begin{figure}
    \includegraphics[width=80mm]{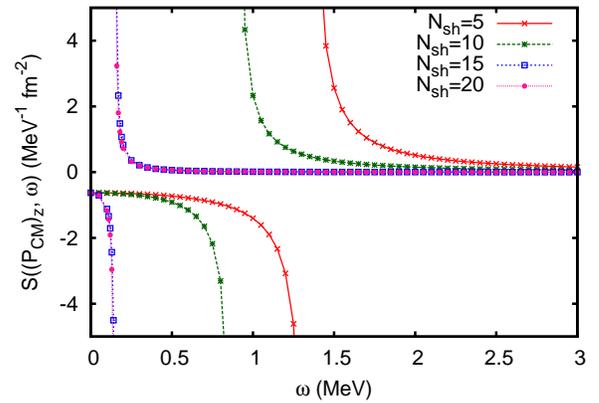}
  \caption{(Color online) The FAM strength function $S((\bm{\hat{P}}_{\rm CM})_z,\omega)$ for a response to the center of mass momentum operator with a real frequency $\omega$.}
  \label{fig:CM-P}
\end{figure}

The center of mass mode appears as a translational symmetry restoration mode.
This is the only case where both the coordinate and momentum operators are known
\begin{align}
  \bm{\hat Q}_{\rm CM} = \frac{1}{A} \sum_{i=1}^A \bm{\hat{r}}_i, \quad
  \bm{\hat P}_{\rm CM} = -i \sum_{i=1}^A \bm{\hat{\nabla}}_i,
\end{align}
and the Thouless-Valatin inertia is nothing but the total mass $M_{\rm CM} = m A$ \cite{Ring-Schuck}.
We use the $z$-component ($K=0$) of the center-of-mass coordinate operator $(\bm{\hat{Q}}_{\rm CM})_z$ as an external field of the FAM.
Figure \ref{fig:CM-Q} shows the FAM strength function with various sizes of the harmonic oscillator single-particle model space $N_{\rm sh}$.
In this calculation a real value of $\omega$ is used, therefore the strength function $S((\bm{\hat{Q}}_{\rm CM})_z,\omega)$ is also real.
As we discussed in Sec.~\ref{sec:approx}, translational symmetry is not an exact symmetry of the system if the single-particle states are expressed in a finite harmonic oscillator basis, and the strength function is approximated by Eq.~(\ref{eq:SQapp}).
Therefore the value of the strength function at $\omega=0$ is finite, and there is a low-energy pole that corresponds to a spurious excitation of the center of mass mode. Its energy approaches zero as the model space size increases.
The strength function for $N_{\rm sh}=20$ is close to that for $N_{\rm sh}=15$, indicating that further convergence of the spurious energy to zero is numerically difficult.
Figure~\ref{fig:CM-P} shows the same plot but for the center-of-mass momentum operator $(\bm{\hat{P}}_{\rm CM})_z$. This curve is approximated by Eq.~(\ref{eq:SPapp}), and the value at $\omega=0$ shows the Thouless-Valatin inertia.

In Table \ref{table:CMmass}, we list the Thouless-Valatin inertia for the center of mass motion in the form of $1/2m=A/2M_{\rm CM}$.
The value for SLy4 is 20.73553 MeV fm$^2$. 
The one-body center-of-mass correction to the kinetic energy \cite{Beiner197529,EPJA7_467,0954-3899-36-10-105105} is not taken into account in this calculation because it effectively scales with the nucleon mass $m$, and makes the comparison complicated.
In a smaller harmonic oscillator model space such as $N_{\rm sh}=5$,
the center of mass excitation has higher excitation energy.
This indicates the coordinate and momentum operators expressed in a small basis are not a good approximation to the NG solution of the QRPA,
and the component of the center of mass can be distributed over the physical modes.
The normalization of the coordinate and momentum operators are satisfied within 0.2\% accuracy.
With a larger harmonic oscillator model space, the spurious energy of the center of mass reaches zero, and the normalization of the operators is more accurate.
In Table \ref{table:CMmass} the Thouless-Valatin inertia computed from the 
energy-weighted sum rule of the center of mass coordinate operator with Eq.~(\ref{eq:TVfromQ}) up to 2 MeV and from the strength function with the center of mass momentum operator at zero energy with Eq.~(\ref{eq:TVfromP}) are listed. Both inertias are close to the exact value, but the inertia computed from the momentum operator agrees precisely, beacuse it is evaluated at a single point $\omega=0$
in the complex-energy plane, and is more precise than the discretized contour integration with the coordinate operator.
Practically, in all cases of symmetry-breaking modes other than the center of mass mode, we know only the momentum operator, and the Thouless-Valatin inertia can be derived from it.
This analysis using the center of mass mode shows that even if the symmetry is approximate, the Thouless-Valatin inertia can be computed very accurately from the FAM strength function.

In the same table, we also show the Inglis-Belyaev inertia for the center of mass motion. As is well known, the Inglis-Belyaev value of the inertia deviates from the Thouless-Valatin inertia, showing that the contribution of the residual interaction is very important even in the case of trivial center of mass motion.

\begin{table*}
  \caption{The inertia for the center of mass motion of $^{26}$Mg in units of MeV fm$^2$. The Thouless-Valatin inertia is computed from the energy weighted sum rule of the center of mass coordinate operator $(\hat{\bm{Q}}_{\rm CM})_z$ with Eq.~(\ref{eq:TVfromQ}) 
    with a radius $R_{A1}=2$ MeV discretized with $N_{A1}=12$ points, and
    from the strength function at zero energy with the center of mass momentum operator $(\hat{\bm{P}}_{\rm CM})_z$ with Eq.~(\ref{eq:TVfromP}).
    The Inglis-Belyaev value of the inertia, the spurious excitation energy of the center of mass mode evaluated with Eq.~(\ref{eq:NGenergy}), and the commutation of the coordinate and momentum operators are also listed.
\label{table:CMmass}}
  \begin{ruledtabular}
  \begin{tabular}{cccccc}
    $N_{\rm sh}$ & $1/2m$ from $(\hat{\bm{Q}}_{\rm CM})_z$ & $1/2m$ from $(\hat{\bm{P}}_{\rm CM})_z$ & Inglis-Belyaev
    & $\Omega_{\rm CM}$ MeV & $\langle[(\hat{\bm{Q}}_{\rm CM})_z,(\hat{\bm{P}}_{\rm CM})_z]\rangle/i$ \\ \hline
    5    & 20.69748 & 20.74676 & 26.04977 & 1.346 & 0.998836 \\
   10    & 20.78073 & 20.82140 & 25.87571 & 0.889 & 0.999310 \\
   15    & 20.73573 & 20.73232 & 25.73650 & 0.151 & 1.000026 \\
   20    & 20.73946 & 20.73666 & 25.74138 & 0.146 & 1.000041 \\ \hline
   exact & 20.73553 & 20.73553 & - & 0 & 1
   \end{tabular}
  \end{ruledtabular}
  \end{table*}


\subsection{Pairing rotational mode}

Unlike center of mass motion,
particle number operators are defined in the configurational space.
Therefore the broken particle-number gauge symmetry is always exact if the same model space is employed in the HFB and QRPA frameworks.

As we have seen in Eq.~(\ref{eq:S_NG_P}), the strength function obtained with the proton particle number field, that is a broken particle-number gauge symmetry in $^{26}$Mg,
$S(\hat{N}_p,\omega)=0$ is zero except at $\omega=0$. The FAM enables us to compute the discontinous value at $\omega=0$ without any convergence problems, and
elsewhere the strength function is numerically zero.
The Thouless-Valatin inertia for the proton pairing rotation in the present case with $N_{\rm sh}=5$ is 1.1545 MeV$^{-1}$. 
The same quantity but for the neutron number operator is exactly zero including at $\omega=0$, because 
the neutron particle number is a conserved symmetry, and commutes with
all other QRPA modes of a particle-hole type.

\begin{figure}
  \includegraphics[width=80mm]{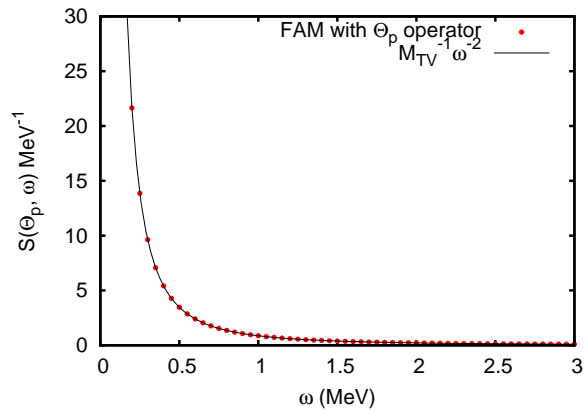}
  \caption{(Color online) The FAM strength function for the proton pairing-rotational angle operator $S(\hat{\Theta}_p,\omega)$ for a real frequency $\omega$ for $^{26}$Mg, computed with a $N_{\rm sh}=5$ model space, plotted together with the curve in Eq.~(\ref{eq:SFwQ}) with the Thouless-Valatin inertia $M_{\rm TV}=$1.1545 MeV$^{-1}$.}
\label{fig:pairrotstrength-Thetap}    
\end{figure}

From Eq.~(\ref{eq:Q_NG}), we can derive the proton gauge angle operator $\hat{\Theta}_p$, which is the conjugate coordinate operator of the proton pairing rotation. Although we can compute the Thouless-Valatin inertia from $S(\hat{N}_p,\omega=0)$, to check the consistency,
we show the response to the coordinate operator of the proton pairing rotation in Fig.~\ref{fig:pairrotstrength-Thetap}. In the figure, the strength function computed from the FAM and Eq.~(\ref{eq:SFwQ}) with the Thouless-Valatin inertia determined from the proton number operator are compared.
Both curves agree very well as expected. We note that this strength function $S(\hat{\Theta}_p,\omega)$ looks very different with that of the center of mass coordinate operator $S((\hat{\bm{Q}}_{\rm CM})_z,\omega)$ in Fig.~\ref{fig:CM-Q}. In the case of the pairing rotational mode, the symmetry is exact, and the position of the pole is exactly at zero energy.
The Thouless-Valatin inertia from the $m_1({\hat{\Theta}_p})$ sum rule through Eq.~(\ref{eq:TVfromQ}) with $R_{A1}=1$ MeV is 1.1545 MeV$^{-1}$, and is perfectly consistent with the inertia from the momentum operator $\hat{N}_p$.

\section{Realistic examples \label{sec:pairrot}}

\subsection{Pairing rotation in single-closed shell}

As for realistic examples of NG modes in nuclei, we discuss pairing rotations in single-closed shell nuclei.
The ground states of tin isotopes are known to form a neutron pairing rotational band
\cite{Ring-Schuck, Broglia20001,Brink-Broglia}, which has been actively discussed recently in connection with the two-neutron transfer reaction \cite{PhysRevC.84.044317,PhysRevC.88.054308,PhysRevLett.107.092501,PhysRevC.87.054321}.
The binding energy of the $N$-neutron isotope $B(N,Z)$ from the nearby reference state with neutron $N_0$ is written as
\begin{align}
  B(N,Z_0) = B(N_0,Z_0) + \lambda_n(N_0,Z_0) \Delta N + \frac{  (\Delta N)^2}{2{\cal J}_n(N_0,Z_0)}, \label{eq:bindingenergy}
\end{align}
where $\Delta N=N-N_0$, and $\lambda_n(N_0,Z_0)$ is the neutron chemical potential of the reference state.
Under the assumption that the ground states of even-even nuclei form
a pairing rotational band, the last term can be regarded as the pairing rotational energy, and the neutron pairing rotational moment of inertia ${\cal J}_n(N_0,Z_0)$ at a reference state is computed with the FAM from the zero-energy response of the neutron particle-number field
\begin{align}
  {\cal J}_n(N_0,Z_0) = -S(\hat{N}_n,\omega=0)
\end{align}
at a reference state.

Again we use SLy4 with volume pairing in a $N_{\rm sh}=20$ model space.
The pairing strength is fixed to
$V_0=-178.81$ MeV fm$^3$ which reproduces the experimental averaged
neutron pairing gap
$\widetilde{\Delta}^{(3)}_n(^{116}{\rm Sn}) = [\Delta^{(3)}_n(^{115}{\rm Sn}) + \Delta_{n}^{(3)}(^{117}{\rm Sn})]/2 =1.100$ MeV \cite{ame2012}.
This parameter setting is used for all the calculations in Sec.~\ref{sec:pairrot}.
We calculated the neutron pairing rotational moment of inertia at the reference state $^{116}$Sn ($N=66$)
that is located in the middle of the $N=50$ and 82 shell gaps.
The Thouless-Valatin moment of inertia for the neutron pairing rotation was
${\cal J}_{n,{\rm TV}}= 5.95$ MeV$^{-1}$,
while the Belyaev inertia was given by ${\cal J}_{n,{\rm IB}}= 4.71$ MeV$^{-1}$, and in this case
the residual interaction contributes to the inertia about 30\%.

\begin{figure}
\includegraphics[width=80mm]{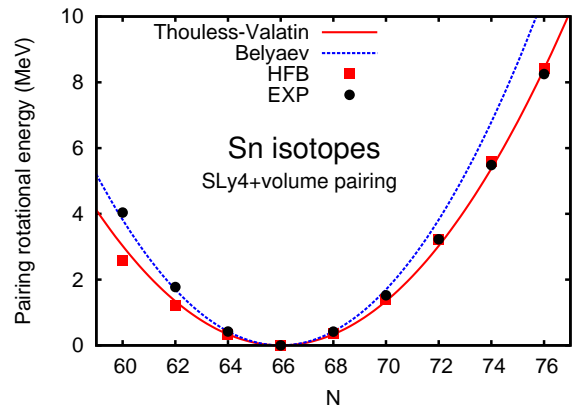}
\caption{ (Color online)
  Neutron pairing rotational energy of even Sn isotopes.
  The red solid (blue dashed) curve is the pairing rotational energy
  $(N-66)^2/[2{\cal J}_{n,{\rm TV(B)}}(^{116}{\rm Sn})]$
  with the Thouless-Valatin (Belyaev) pairing rotational moment of inertia evaluated at $^{116}$Sn.
  The red squares are the HFB energy $E_{\rm HFB}(N) - E_{\rm HFB}(^{116}{\rm Sn}) - \lambda_{n,{\rm HFB}}(^{116}{\rm Sn})(N-66)$,
  and the black circles are the experimental values evaluated with $-B_{\rm exp}(N) - \lambda_{n,{\rm exp}}(^{116}{\rm Sn})(N-66)$, where
  the binding energy is taken from Ref.~\cite{ame2012}.
  The experimental neutron chemical potential is evaluated with
 $\lambda_{n, {\rm exp}}(^{116}{\rm Sn})
= [B_{\rm exp}(^{118}{\rm Sn}) - B_{\rm exp}(^{114}{\rm Sn})]/4=-8.345$ MeV.
  \label{fig:Sn_pairrot}}
\end{figure}

Figure~\ref{fig:Sn_pairrot} shows the neutron pairing rotational energy
measured from $^{116}$Sn as a function of neutron number. 
The Thouless-Valatin inertia for the neutron pairing rotation explains the pairing rotational spectrum especially well in the vicinity of the reference state.
Agreement between the pairing rotational energy from the Thouless-Valatin inertia and the HFB energy shows the validity of the dielectric theorem far from the reference state and at small anharmonicity.
We note that inclusion of the one-body center-of-mass correction violates the dielectric theorem, because the correction term $(1-1/A)$ is not variational with respect to the change of particle numbers \cite{PhysRevC.77.024316}.
As the neutron number changes from $N=66$, deviation from the parabola curve is seen, both in the HFB calculations and the experimental data, indicating a change of the intrinsic structures as a function of the neutron number. The deviation is larger in the neutron-deficient side.

\begin{figure}
\includegraphics[width=80mm]{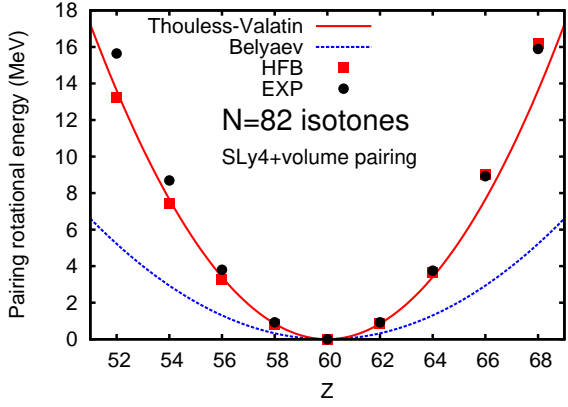}
\caption{ (Color online)
  Same as Fig.~\ref{fig:Sn_pairrot} but for the proton pairing rotational energy for even $N=82$ isotones measured from $^{142}$Nd $(Z=60)$. The experimental proton chemical potential is $\lambda_{p,{\rm exp}}(^{142}{\rm Nd}) = -5.761$ MeV.
  \label{fig:N82_pairrot}
}
\end{figure}

The next example is the proton pairing rotation in $N=82$ isotones,
where we take $^{142}$Nd ($Z=60$) as our reference state.
The averaged proton pairing gap $\widetilde{\Delta}^{(3)}_p(^{142}{\rm Nd})=0.788$ MeV is well reproduced with the same pairing functional parameters.
We can define the proton pairing rotational moment of inertia ${\cal J}_p(N_0,Z_0)$
by repeating the same discussion with Eq.~(\ref{eq:bindingenergy}) but for protons.
The Thouless-Valatin inertia for this proton pairing rotation is 2.35 MeV$^{-1}$,
while the Belyaev inertia is 6.13 MeV$^{-1}$. The residual interaction changes the pairing rotational moment of inertia by a factor of about 2.6.
This is because of the residual Coulomb contribution:
the Coulomb interaction is known to affect the proton pairing energy and gap at the mean-field level \cite{Anguiano2001227,EPJA40_121,PhysRevC.83.031302}.
Because the Coulomb energy is proportional to $Z^2$,
its residual part directly contributes to the proton pairing rotational moment of inertia in the QRPA level.
Figure~\ref{fig:N82_pairrot} shows the proton pairing rotational energy measured from $^{142}$Nd. To explain the experimental curve, the contribution of the residual interaction is essential in the case of the proton pairing rotation.

\subsection{Mixing of neutron and proton pairing rotational modes}

When both the neutron and proton are in the superconducting phase, the neutron pairing and proton pairing rotational modes appear as NG modes.
These zero-energy modes are degenerate because they have the same $K^\pi=0^+$ quantum numbers. Thus the eigenmodes as solutions of the QRPA equations are generally the linear combination of the two pairing rotational modes. The momentum operators of the two NG modes are written as
\begin{align}
  \hat{N}_1 =& \hat{N}_n \cos\theta + \alpha \hat{N}_p \sin\theta, \nonumber \\
  \hat{N}_2 =& -\hat{N}_n \sin\theta + \alpha \hat{N}_p \cos\theta,  \label{eq:N}
\end{align}
where $\theta$ is a mixing angle, and $\alpha$ is a scaling parameter.
The parameter $\alpha$ should be one under the isospin symmetry, but we keep a general expression here. The overall scaling of the operators $\hat{N}_1$ and $\hat{N}_2$ does not change the physics.
The conjugate angle operators are written as
\begin{align}
  \hat{\Theta}_1 =& \hat{\Theta}_n \cos \theta + \frac{1}{\alpha} \hat{\Theta}_p \sin\theta \nonumber  \\
  \hat{\Theta}_2 =& -\hat{\Theta}_n \sin\theta + \frac{1}{\alpha}\hat{\Theta}_p \cos\theta, \label{eq:Theta}
\end{align}
where $\hat{\Theta}_n$ and $\hat{\Theta}_p$ are the neutron and proton two-quasiparticle parts of the operator. The operators in Eqs.~(\ref{eq:N}) and (\ref{eq:Theta}) satisfy the commutation relations of Eq.~(\ref{eq:PQnorm}) if the neutron and proton parts of the operators are normalized with the same commutation relation.

The Thouless-Valatin inertias for the two NG modes are derived from Eq.~(\ref{eq:TVMOI-QRPA}):
\begin{align}
  M_1 =& -S(\hat{N}_n,\hat{N}_n) \cos^2\theta -\alpha^2S(\hat{N}_p,\hat{N}_p)\sin^2\theta \nonumber \\
   &- 2\alpha S(\hat{N}_n,\hat{N}_p)\sin\theta\cos\theta, \\
  M_2 =& -S(\hat{N}_n,\hat{N}_n) \sin^2\theta -\alpha^2S(\hat{N}_p,\hat{N}_p)\cos^2\theta \nonumber \\
  & + 2\alpha S(\hat{N}_n,\hat{N}_p)\sin\theta\cos\theta, 
\end{align}
where
\begin{align}
  S(\hat{N}_n,\hat{N}_n) =& -2 N_n(A+B)^{-1}N_n = S(\hat{N}_n,\omega=0),\\
  S(\hat{N}_n,\hat{N}_p) =& -2 N_n(A+B)^{-1}N_p = S(\hat{N}_p,\hat{N}_n),\\
  S(\hat{N}_p,\hat{N}_p) =& -2 N_p(A+B)^{-1}N_p = S(\hat{N}_p,\omega=0)
\end{align}
are the strength functions obtained from the FAM with zero frequencies.
Here $N_n$ and $N_p$ are the two-quasiparticle amplitudes of the particle number operators $\hat{N}_n$ and $\hat{N}_p$, respectively (we assume ${\rm Im}\,\, N_n = {\rm Im}\,\, N_p = 0$).
The off-diagonal term $S(\hat{N}_n,\hat{N}_p)$ is obtained by evaluating the strength function for the proton particle number operator from the FAM equations with a neutron particle-number external field, or vice versa.
The two parameters $\theta$ and $\alpha$ are constrained  from the orthogonality of the two NG modes
\begin{align}
  \tan 2\theta = \frac{ 2\alpha S(\hat{N}_n,\hat{N}_p)} {S(\hat{N}_n,\hat{N}_n) - \alpha^2 S(\hat{N}_p,\hat{N}_p)}.
\end{align}
In comparison with experimental data, we are interested in the pairing rotational energy in terms of neutrons and protons rather than in terms of the neutron-proton mixed eigenmodes. The pairing rotational energy is written as
\begin{align}
  E_{\rm rot}(N,Z) =& \frac{(\Delta N_1)^2}{2M_1} + \frac{(\Delta N_2)^2}{2M_2} \nonumber \\
  =& \frac{1}{2}\begin{pmatrix} \Delta N & \Delta Z \end{pmatrix}
  {\mathbb J}^{-1}
  \begin{pmatrix} \Delta N \\ \Delta Z \end{pmatrix} \nonumber \\
  =& \frac{ (\Delta N)^2}{2{\cal J}_{nn}} +
  \frac{2(\Delta N) (\Delta Z)}{2{\cal J}_{np}}  +
  \frac{(\Delta Z)^2 }{2{\cal J}_{pp}},   \label{eq:Epair}
\end{align}
where $\Delta N_1, \Delta Z_1, \Delta N$ and $\Delta Z$ are the deviation of the particle numbers from a reference state.
The inertia tensor ${\mathbb J}$  is given by
\begin{align}
  {\mathbb J}^{-1} = 
  -\begin{pmatrix} S(\hat{N}_n,\hat{N}_n) & S(\hat{N}_n,\hat{N}_p) \\
    S(\hat{N}_p,\hat{N}_n) & S(\hat{N}_p,\hat{N}_p) \end{pmatrix}^{-1}
  = \begin{pmatrix}
    1/{\cal J}_{nn} & 1/{\cal J}_{np} \\ 1/{\cal J}_{pn} &     1/{\cal J}_{pp}
    \end{pmatrix},
\end{align}
and the neutron and proton components are explicitly written as
\begin{align}
  {\cal J}_{nn} =& -\frac{ S(\hat{N}_n,\hat{N}_n)S(\hat{N}_p,\hat{N}_p)-S(\hat{N}_n,\hat{N}_p)^2}{S(\hat{N}_p,\hat{N}_p)}, \\
  {\cal J}_{np} =& \frac{ S(\hat{N}_n,\hat{N}_n)S(\hat{N}_p,\hat{N}_p)-S(\hat{N}_n,\hat{N}_p)^2}{S(\hat{N}_n,\hat{N}_p)}, \\
    {\cal J}_{pp} =& -\frac{ S(\hat{N}_n,\hat{N}_n)S(\hat{N}_p,\hat{N}_p)-S(\hat{N}_n,\hat{N}_p)^2}{S(\hat{N}_n,\hat{N}_n)}.  
\end{align}
This shows that the principal axes of the pairing rotations are not aligned with
the neutron and proton directions in the gauge space
because of the presence of the off-diagonal term $S(\hat{N}_n,\hat{N}_p)$.
We note that this does not exist in the Belyaev inertia, because the two-quasiparticle indices in Eq.~(\ref{eq:cranking}) are either neutrons or protons when neutron-proton mixing is absent in the mean field. The residual interaction plays an essential role for generating the neutron-proton term in the pairing rotational energy.
In Ref.~\cite{PhysRevC.90.014312}, the principal axes tilted against the neutron and proton gauge-angle space have been reported
in the reduced energy kernel when neutrons and protons are superconducting.

\subsection{Neutron and proton pairing rotations around $^{130}$Xe}

We consider the neutron and proton pairing rotations by taking a reference state at the open-shell deformed nucleus $^{130}$Xe.
The lowest energy HFB solution obtained with the axial HFB code has a prolate deformation with $\beta=0.143$, and both neutrons and protons are superconducting with $\Delta_n = 0.702$ MeV and $\Delta_p=0.517$ MeV.
The experimental averaged pairing gaps are
$\widetilde{\Delta}^{(3)}_n(^{130}{\rm Xe})=1.170$ MeV and 
$\widetilde{\Delta}^{(3)}_p(^{130}{\rm Xe})=1.014$ MeV.

The response functions computed from the FAM are $S(\hat{N}_n,\hat{N}_n) = 12.704$ MeV$^{-1}$,
$S(\hat{N}_n,\hat{N}_p) = 8.725$ MeV$^{-1}$, and $S(\hat{N}_p,\hat{N}_p) = 3.083$ MeV$^{-1}$. We took a numerical average of $S(\hat{N}_n,\hat{N}_p)$ and $S(\hat{N}_p,\hat{N}_n)$ for the off-diagonal term.
The Thouless-Valatin moments of inertia are ${\cal J}_{nn} = 11.986$ MeV$^{-1}$,
${\cal J}_{np}= -4.236$ MeV$^{-1}$, and ${\cal J}_{pp}=2.909$ MeV$^{-1}$.
The opposite sign of ${\cal J}_{np}$ is consistent with the isorotation picture, whose rotational energy is proportional to $T(T+1)$ \cite{PhysRevLett.23.1299,Dussel1970469,PhysRevC.80.044313}, and produces a negative sign for the neutron-proton term.

\begin{figure*}
    \begin{tabular}{cc}
      \includegraphics[width=80mm]{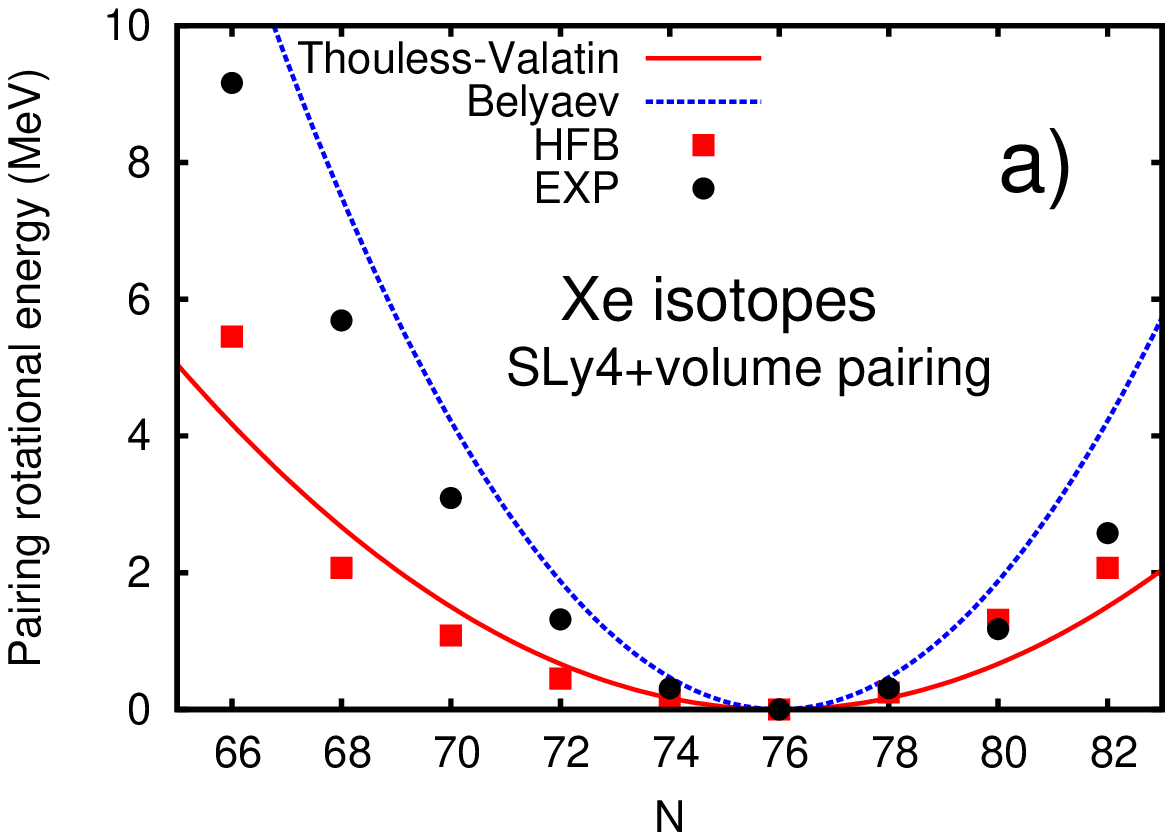} & 
      \includegraphics[width=80mm]{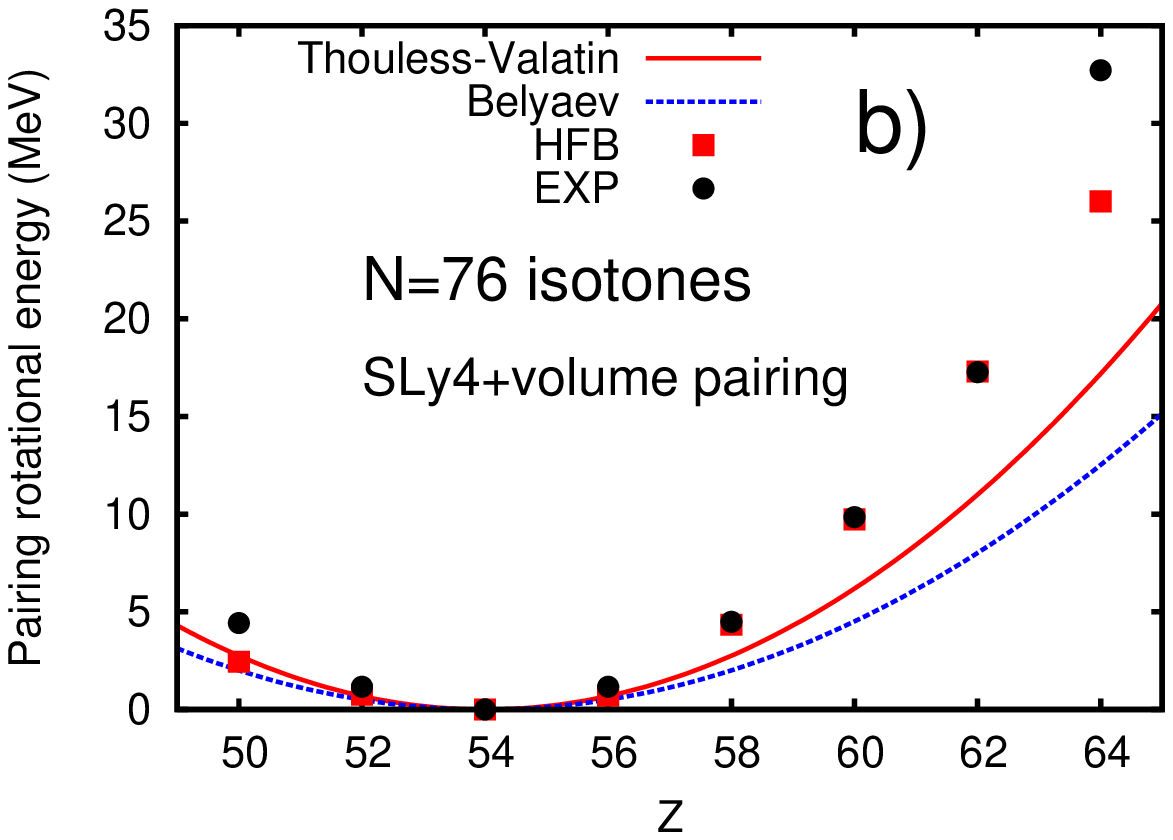}\\
      \includegraphics[width=80mm]{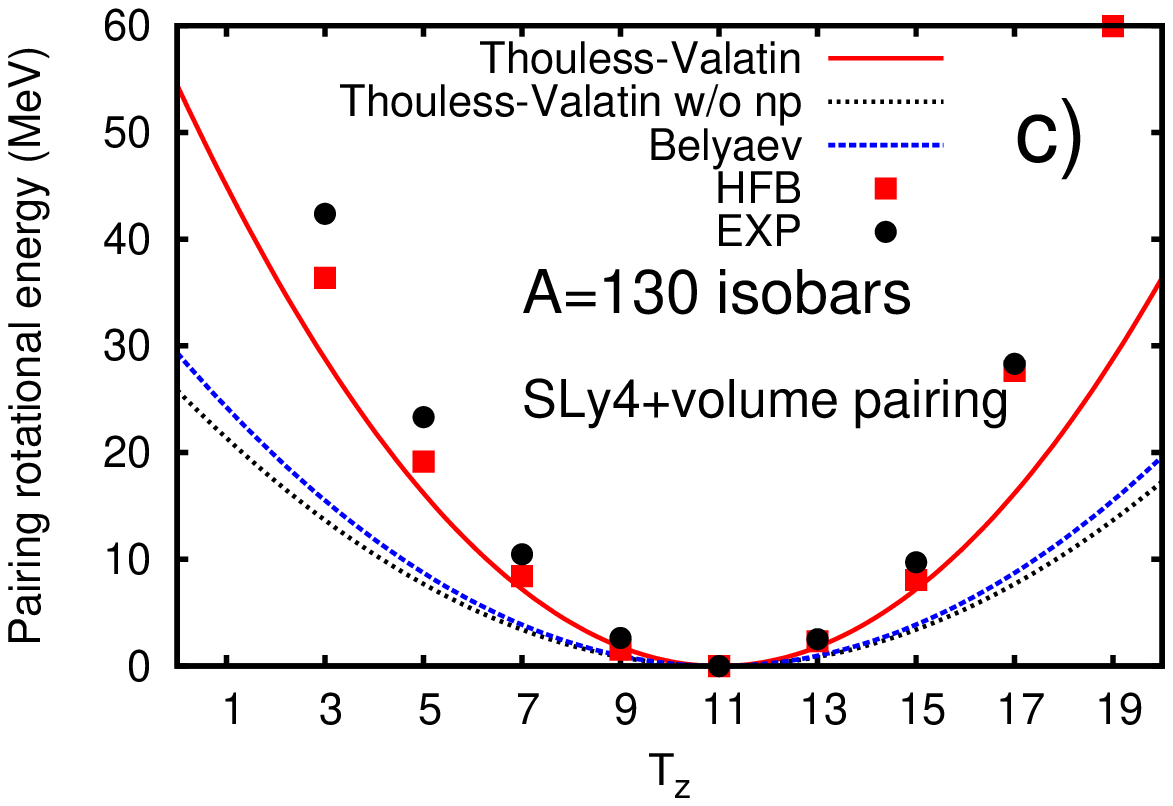} &
      \includegraphics[width=80mm]{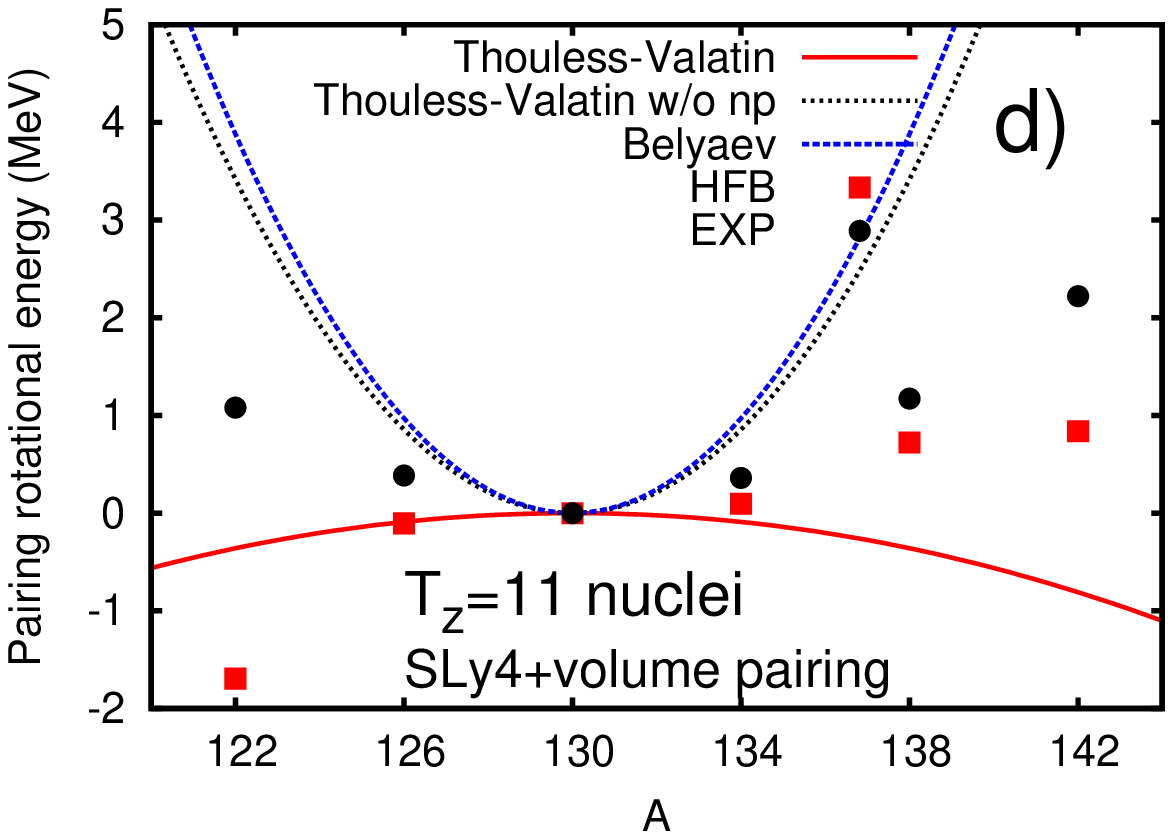}    
    \end{tabular}
    \caption{ (Color online) 
      a) Neutron pairing rotational energy along Xe isotopes,
      b) proton pairing rotational energy along $N=76$ isotones,
      c) neutron and proton pairing rotational energies along the $A=130$ isobar, and
      d) along $T_z=11$ nuclei
      measured from $^{130}$Xe ($N=76,Z=54$).
      The experimental neutron and proton chemical potentials are
      $\lambda_{n,{\rm exp}}=-7.926$ MeV  and $\lambda_{p,{\rm exp}} =-7.149$ MeV.
      The red solid curves are the pairing rotational energies with the Thouless-Valatin inertias in Eq.~(\ref{eq:Epair}), black dotted curves are the same energies but without the neutron-proton terms. The blue dashed curves are the pairing rotational energies from the Belyaev inertias without neutron-proton terms.
      \label{fig:130Xe} }
  \end{figure*}

Figure \ref{fig:130Xe} shows the pairing rotational energies measured from $^{130}$Xe along the Xe isotope direction, the $N=76$ isotone direction, the $A=130$ isobar direction, and the $T_z=11$ direction.
Clear parabola patterns are seen in the pairing rotational energy from the HFB and the experimental data in Fig.~\ref{fig:130Xe} a) and b), although the reference state has a prolate deformation and the intrinsic shape changes nucleus by nucleus.
The Thouless-Valatin inertia explains the neutron and proton pairing rotational energy in the vicinity of the reference state.

In Fig.~\ref{fig:130Xe} c), the pairing rotational energy along the $A=130$ isobar is shown. Again a parabola pattern is found along the isobar, and both the HFB and the Thouless-Valatin inertia explain the experimental data.
From Eq.~(\ref{eq:Epair}) the pairing rotational energy along the isobar is given by
\begin{align}
    E_{\rm rot}(N,Z) = \left(
  \frac{1}{2{\cal J}_{nn}} - \frac{2}{2{\cal J}_{np}} + \frac{1}{2{\cal J}_{pp}}
  \right) (\Delta T_z)^2
\end{align}
with $\Delta T_z = 11 - T_z$.
This represents the isorotational energy which restores the broken isospin symmetry. The coefficient in front of $(\Delta T_z)^2$ is 0.44 MeV, and explains the systematic behavior of the binding energies.
For comparison we also show  the pairing rotational energy with the Thouless-Valatin inertia but without the neutron-proton term in the same figure. The value is close to the Belyaev inertia that does not contain the neutron-proton term either, indicating the importance of the neutron-proton term for the isorotation.

The last example is the $T_z=11$ nuclei shown in Fig.~\ref{fig:130Xe} d).
All the calculations and the experimental data show the pairing rotational energy is small.
This degree of freedom is associated with the breaking of global gauge invariance (the total particle number symmetry) \cite{PhysRevLett.23.1299}.
In fact, the pairing rotational energy from the Thouless-Valatin inertia along a constant $T_z$ line is written as
\begin{align}
  E_{\rm rot}(N,Z) = \left(
  \frac{1}{2{\cal J}_{nn}} + \frac{2}{2{\cal J}_{np}} + \frac{1}{2{\cal J}_{pp}}
  \right) (\Delta A)^2,
\end{align}
and the coefficient in front of $(\Delta A)^2=(A-130)^2$ is $-0.02$ MeV.
Our functional preserves an approximate isospin symmetry well, and 
only the Coulomb term breaks it.
Using independent pairing strengths for neutrons and protons introduces
explicit isospin symmetry breaking in the pairing channel, and may generate pairing rotational energy associated with global gauge invariance.
The quadrupole shape changes rapidly along the $T_z=11$ chain,
and the correlation energy from the deformation is larger than the pairing rotational energy. This causes a deviation from the parabola curve in the pairing rotational energy.

As a whole, the pairing rotational description based on the open shell nucleus $^{130}$Xe works well, and explains the experimental binding energy systematics of neighboring even-even nuclei around $^{130}$Xe. This indicates that the intrinsic superconducting HFB state of $^{130}$Xe contains information of neighboring even-even nuclei.
With this generalization of the pairing rotation to the neutron and proton mixed modes, we can take an arbitrary superconducting nucleus as a reference state of the pairing rotation. Systematic analysis of the pairing rotational modes and moments of inertia in this direction is in progress \cite{nhinprep}.

\section{Conclusion \label{sec:conclusion}}

We formulated linear response theory in the presence of zero-energy NG modes. We showed that the Thouless-Valatin inertia of the NG mode is derived from the zero-frequency response to the momentum operator of the NG mode. Combined with the finite-amplitude method for nuclear energy density functional theory, we can compute the Thouless-Valatin inertia very precisely and efficiently. This formulation also provides the expression for the conjugate coordinate operator that will be necessary for the spurious mode removal procedure of the FAM.
The formulation was numerically tested in detail for the cases of the trivial center of mass mode and the pairing rotational mode. Although the center of mass mode is not at zero energy in practical numerical calculations, the strength function at zero frequency provides a precise Thouless-Valatin inertia.
The realistic applications for the neutron and proton pairing rotational moments of inertia are presented for Sn isotopes and $N=82$ isotones, respectively.
We then consider the situation where both neutrons and protons are in a superconducting phase. We show that the neutron and proton pairing rotational degrees of freedom are mixed in the QRPA normal modes, and the neutron-proton term in the pairing rotational energy is generated. As a realistic application, we show that the ground state energies around $^{130}$Xe can be interpreted in terms of the pairing rotation picture.

One interesting future application of this formalism for NG modes is in the computation of the Thouless-Valatin rotational moment of inertia.
A systematic comparison of the energy of the $2_1^+$ states of deformed nuclei with experimental data may clarify the property of the unconstrained time-odd term  of the nuclear EDF.
Once implementation into the symmetry unrestricted HFB code has been completed, this approach can serve as an efficient technique for deriving the Thouless-Valatin moments of inertia of three-dimensional collective rotation for the five-dimensional quadrupole collective Hamiltonian \cite{PhysRevC.82.064313}.

Deriving the pairing collective Hamiltonian \cite{Bes19701,PhysRevLett.96.032501,PhysRevC.77.057301} based on the nuclear EDF and discussing the anharmonic and large-amplitude aspect of the collective pairing motion and the coupling of the pairing vibration and pairing rotation will be challenges for the future.
Extension of the neutron and proton pairing rotations to full $SU(2)$ isorotation by including the neutron-proton pairing within the isospin-invariant nuclear density functional theory \cite{PhysRevC.69.014316,PhysRevC.88.061301,PhysRevC.89.054317} is another future challenge for understanding the role of the isospin symmetry in low-energy nuclear collective motion.

\section*{Acknowledgments}

The author thanks W. Nazarewicz and E. Olsen for their careful reading of the manuscript and constructive comments.
Useful discussions with G. Col\`{o}, M. Kortelainen and T. Oishi are gratefully acknowledged.
Numerical calculation was performed in the resources of High Performance Computing Center, Institute for Cyber-Enabled Research, Michigan State University, and the COMA (PACS-IX) System at the Center for Computational Sciences, University of Tsukuba.

\bibliographystyle{apsrev4-1}

\begin{thebibliography}{73}%
\makeatletter
\providecommand \@ifxundefined [1]{%
 \@ifx{#1\undefined}
}%
\providecommand \@ifnum [1]{%
 \ifnum #1\expandafter \@firstoftwo
 \else \expandafter \@secondoftwo
 \fi
}%
\providecommand \@ifx [1]{%
 \ifx #1\expandafter \@firstoftwo
 \else \expandafter \@secondoftwo
 \fi
}%
\providecommand \natexlab [1]{#1}%
\providecommand \enquote  [1]{``#1''}%
\providecommand \bibnamefont  [1]{#1}%
\providecommand \bibfnamefont [1]{#1}%
\providecommand \citenamefont [1]{#1}%
\providecommand \href@noop [0]{\@secondoftwo}%
\providecommand \href [0]{\begingroup \@sanitize@url \@href}%
\providecommand \@href[1]{\@@startlink{#1}\@@href}%
\providecommand \@@href[1]{\endgroup#1\@@endlink}%
\providecommand \@sanitize@url [0]{\catcode `\\12\catcode `\$12\catcode
  `\&12\catcode `\#12\catcode `\^12\catcode `\_12\catcode `\%12\relax}%
\providecommand \@@startlink[1]{}%
\providecommand \@@endlink[0]{}%
\providecommand \url  [0]{\begingroup\@sanitize@url \@url }%
\providecommand \@url [1]{\endgroup\@href {#1}{\urlprefix }}%
\providecommand \urlprefix  [0]{URL }%
\providecommand \Eprint [0]{\href }%
\providecommand \doibase [0]{http://dx.doi.org/}%
\providecommand \selectlanguage [0]{\@gobble}%
\providecommand \bibinfo  [0]{\@secondoftwo}%
\providecommand \bibfield  [0]{\@secondoftwo}%
\providecommand \translation [1]{[#1]}%
\providecommand \BibitemOpen [0]{}%
\providecommand \bibitemStop [0]{}%
\providecommand \bibitemNoStop [0]{.\EOS\space}%
\providecommand \EOS [0]{\spacefactor3000\relax}%
\providecommand \BibitemShut  [1]{\csname bibitem#1\endcsname}%
\let\auto@bib@innerbib\@empty
\bibitem [{\citenamefont {Nazarewicz}(1994)}]{Nazarewicz199427}%
  \BibitemOpen
  \bibfield  {author} {\bibinfo {author} {\bibfnamefont {W.}~\bibnamefont
  {Nazarewicz}},\ }\href {\doibase
  http://dx.doi.org/10.1016/0375-9474(94)90037-X} {\bibfield  {journal}
  {\bibinfo  {journal} {Nucl. Phys. A}\ }\textbf {\bibinfo {volume} {574}},\
  \bibinfo {pages} {27 } (\bibinfo {year} {1994})}\BibitemShut {NoStop}%
\bibitem [{\citenamefont {Frauendorf}(2001)}]{RevModPhys.73.463}%
  \BibitemOpen
  \bibfield  {author} {\bibinfo {author} {\bibfnamefont {S.}~\bibnamefont
  {Frauendorf}},\ }\href {\doibase 10.1103/RevModPhys.73.463} {\bibfield
  {journal} {\bibinfo  {journal} {Rev. Mod. Phys.}\ }\textbf {\bibinfo {volume}
  {73}},\ \bibinfo {pages} {463} (\bibinfo {year} {2001})}\BibitemShut
  {NoStop}%
\bibitem [{\citenamefont {Satu\l{}a}\ and\ \citenamefont
  {Wyss}(2005)}]{0034-4885-68-1-R03}%
  \BibitemOpen
  \bibfield  {author} {\bibinfo {author} {\bibfnamefont {W.}~\bibnamefont
  {Satu\l{}a}}\ and\ \bibinfo {author} {\bibfnamefont {R.~A.}\ \bibnamefont
  {Wyss}},\ }\href {http://stacks.iop.org/0034-4885/68/i=1/a=R03} {\bibfield
  {journal} {\bibinfo  {journal} {Rep. Prog. Phys.}\ }\textbf {\bibinfo
  {volume} {68}},\ \bibinfo {pages} {131} (\bibinfo {year} {2005})}\BibitemShut
  {NoStop}%
\bibitem [{\citenamefont {Yannouleas}\ and\ \citenamefont
  {Landman}(2007)}]{0034-4885-70-12-R02}%
  \BibitemOpen
  \bibfield  {author} {\bibinfo {author} {\bibfnamefont {C.}~\bibnamefont
  {Yannouleas}}\ and\ \bibinfo {author} {\bibfnamefont {U.}~\bibnamefont
  {Landman}},\ }\href {http://stacks.iop.org/0034-4885/70/i=12/a=R02}
  {\bibfield  {journal} {\bibinfo  {journal} {Rep. Prog. Phys.}\ }\textbf
  {\bibinfo {volume} {70}},\ \bibinfo {pages} {2067} (\bibinfo {year}
  {2007})}\BibitemShut {NoStop}%
\bibitem [{\citenamefont {Birman}\ \emph {et~al.}(2013)\citenamefont {Birman},
  \citenamefont {Nazmitdinov},\ and\ \citenamefont {Yukalov}}]{Birman20131}%
  \BibitemOpen
  \bibfield  {author} {\bibinfo {author} {\bibfnamefont {J.}~\bibnamefont
  {Birman}}, \bibinfo {author} {\bibfnamefont {R.}~\bibnamefont {Nazmitdinov}},
  \ and\ \bibinfo {author} {\bibfnamefont {V.}~\bibnamefont {Yukalov}},\ }\href
  {\doibase http://dx.doi.org/10.1016/j.physrep.2012.11.005} {\bibfield
  {journal} {\bibinfo  {journal} {Phys. Rep.}\ }\textbf {\bibinfo {volume}
  {526}},\ \bibinfo {pages} {1 } (\bibinfo {year} {2013})}\BibitemShut
  {NoStop}%
\bibitem [{\citenamefont {Frauendorf}\ and\ \citenamefont
  {Sheikh}(2000)}]{1402-4896-2000-T88-032}%
  \BibitemOpen
  \bibfield  {author} {\bibinfo {author} {\bibfnamefont {S.}~\bibnamefont
  {Frauendorf}}\ and\ \bibinfo {author} {\bibfnamefont {J.~A.}\ \bibnamefont
  {Sheikh}},\ }\href {http://stacks.iop.org/1402-4896/T88/162} {\bibfield
  {journal} {\bibinfo  {journal} {Phys. Scr.}\ }\textbf {\bibinfo {volume}
  {T88}},\ \bibinfo {pages} {162} (\bibinfo {year} {2000})}\BibitemShut
  {NoStop}%
\bibitem [{\citenamefont {Neerg\aa{}rd}(2009)}]{PhysRevC.80.044313}%
  \BibitemOpen
  \bibfield  {author} {\bibinfo {author} {\bibfnamefont {K.}~\bibnamefont
  {Neerg\aa{}rd}},\ }\href {\doibase 10.1103/PhysRevC.80.044313} {\bibfield
  {journal} {\bibinfo  {journal} {Phys. Rev. C}\ }\textbf {\bibinfo {volume}
  {80}},\ \bibinfo {pages} {044313} (\bibinfo {year} {2009})}\BibitemShut
  {NoStop}%
\bibitem [{\citenamefont {Satu\l{}a}\ \emph {et~al.}(2011)\citenamefont
  {Satu\l{}a}, \citenamefont {Dobaczewski}, \citenamefont {Nazarewicz},\ and\
  \citenamefont {Rafalski}}]{PhysRevLett.106.132502}%
  \BibitemOpen
  \bibfield  {author} {\bibinfo {author} {\bibfnamefont {W.}~\bibnamefont
  {Satu\l{}a}}, \bibinfo {author} {\bibfnamefont {J.}~\bibnamefont
  {Dobaczewski}}, \bibinfo {author} {\bibfnamefont {W.}~\bibnamefont
  {Nazarewicz}}, \ and\ \bibinfo {author} {\bibfnamefont {M.}~\bibnamefont
  {Rafalski}},\ }\href {\doibase 10.1103/PhysRevLett.106.132502} {\bibfield
  {journal} {\bibinfo  {journal} {Phys. Rev. Lett.}\ }\textbf {\bibinfo
  {volume} {106}},\ \bibinfo {pages} {132502} (\bibinfo {year}
  {2011})}\BibitemShut {NoStop}%
\bibitem [{\citenamefont {Nambu}(1960)}]{PhysRev.117.648}%
  \BibitemOpen
  \bibfield  {author} {\bibinfo {author} {\bibfnamefont {Y.}~\bibnamefont
  {Nambu}},\ }\href {\doibase 10.1103/PhysRev.117.648} {\bibfield  {journal}
  {\bibinfo  {journal} {Phys. Rev.}\ }\textbf {\bibinfo {volume} {117}},\
  \bibinfo {pages} {648} (\bibinfo {year} {1960})}\BibitemShut {NoStop}%
\bibitem [{\citenamefont {Goldstone}(1961)}]{INC_19_154}%
  \BibitemOpen
  \bibfield  {author} {\bibinfo {author} {\bibfnamefont {J.}~\bibnamefont
  {Goldstone}},\ }\href {\doibase 10.1007/BF02812722} {\bibfield  {journal}
  {\bibinfo  {journal} {Il Nuovo Cimento}\ }\textbf {\bibinfo {volume} {19}},\
  \bibinfo {pages} {154} (\bibinfo {year} {1961})}\BibitemShut {NoStop}%
\bibitem [{\citenamefont {Ring}\ and\ \citenamefont
  {Schuck}(1980)}]{Ring-Schuck}%
  \BibitemOpen
  \bibfield  {author} {\bibinfo {author} {\bibfnamefont {P.}~\bibnamefont
  {Ring}}\ and\ \bibinfo {author} {\bibfnamefont {P.}~\bibnamefont {Schuck}},\
  }\href@noop {} {\emph {\bibinfo {title} {The Nuclear Many-Body Problem}}}\
  (\bibinfo  {publisher} {Springer-Verlag},\ \bibinfo {year}
  {1980})\BibitemShut {NoStop}%
\bibitem [{\citenamefont {Thouless}\ and\ \citenamefont
  {Valatin}(1962)}]{Thouless1962211}%
  \BibitemOpen
  \bibfield  {author} {\bibinfo {author} {\bibfnamefont {D.~J.}\ \bibnamefont
  {Thouless}}\ and\ \bibinfo {author} {\bibfnamefont {J.~G.}\ \bibnamefont
  {Valatin}},\ }\href {\doibase DOI: 10.1016/0029-5582(62)90741-1} {\bibfield
  {journal} {\bibinfo  {journal} {Nucl. Phys.}\ }\textbf {\bibinfo {volume}
  {31}},\ \bibinfo {pages} {211 } (\bibinfo {year} {1962})}\BibitemShut
  {NoStop}%
\bibitem [{\citenamefont {Kammuri}(1967)}]{Kammuri01061967}%
  \BibitemOpen
  \bibfield  {author} {\bibinfo {author} {\bibfnamefont {T.}~\bibnamefont
  {Kammuri}},\ }\href {\doibase 10.1143/PTP.37.1131} {\bibfield  {journal}
  {\bibinfo  {journal} {Prog. Theor. Phys.}\ }\textbf {\bibinfo {volume}
  {37}},\ \bibinfo {pages} {1131} (\bibinfo {year} {1967})}\BibitemShut
  {NoStop}%
\bibitem [{\citenamefont {Bertsch}\ and\ \citenamefont
  {Hagino}(2001)}]{PAN64_588}%
  \BibitemOpen
  \bibfield  {author} {\bibinfo {author} {\bibfnamefont {G.}~\bibnamefont
  {Bertsch}}\ and\ \bibinfo {author} {\bibfnamefont {K.}~\bibnamefont
  {Hagino}},\ }\href {\doibase 10.1134/1.1368217} {\bibfield  {journal}
  {\bibinfo  {journal} {Physics of Atomic Nuclei}\ }\textbf {\bibinfo {volume}
  {64}},\ \bibinfo {pages} {588} (\bibinfo {year} {2001})}\BibitemShut
  {NoStop}%
\bibitem [{\citenamefont {Pr\'{o}chniak}\ and\ \citenamefont
  {Rohozi\'{n}ski}(2009)}]{0954-3899-36-12-123101}%
  \BibitemOpen
  \bibfield  {author} {\bibinfo {author} {\bibfnamefont {L.}~\bibnamefont
  {Pr\'{o}chniak}}\ and\ \bibinfo {author} {\bibfnamefont {S.~G.}\ \bibnamefont
  {Rohozi\'{n}ski}},\ }\href {http://stacks.iop.org/0954-3899/36/123101}
  {\bibfield  {journal} {\bibinfo  {journal} {J. Phys. G}\ }\textbf {\bibinfo
  {volume} {36}},\ \bibinfo {pages} {123101} (\bibinfo {year}
  {2009})}\BibitemShut {NoStop}%
\bibitem [{\citenamefont {Hinohara}\ \emph {et~al.}(2010)\citenamefont
  {Hinohara}, \citenamefont {Sato}, \citenamefont {Nakatsukasa}, \citenamefont
  {Matsuo},\ and\ \citenamefont {Matsuyanagi}}]{PhysRevC.82.064313}%
  \BibitemOpen
  \bibfield  {author} {\bibinfo {author} {\bibfnamefont {N.}~\bibnamefont
  {Hinohara}}, \bibinfo {author} {\bibfnamefont {K.}~\bibnamefont {Sato}},
  \bibinfo {author} {\bibfnamefont {T.}~\bibnamefont {Nakatsukasa}}, \bibinfo
  {author} {\bibfnamefont {M.}~\bibnamefont {Matsuo}}, \ and\ \bibinfo {author}
  {\bibfnamefont {K.}~\bibnamefont {Matsuyanagi}},\ }\href {\doibase
  10.1103/PhysRevC.82.064313} {\bibfield  {journal} {\bibinfo  {journal} {Phys.
  Rev. C}\ }\textbf {\bibinfo {volume} {82}},\ \bibinfo {pages} {064313}
  (\bibinfo {year} {2010})}\BibitemShut {NoStop}%
\bibitem [{\citenamefont {Inglis}(1956)}]{PhysRev.103.1786}%
  \BibitemOpen
  \bibfield  {author} {\bibinfo {author} {\bibfnamefont {D.~R.}\ \bibnamefont
  {Inglis}},\ }\href {\doibase 10.1103/PhysRev.103.1786} {\bibfield  {journal}
  {\bibinfo  {journal} {Phys. Rev.}\ }\textbf {\bibinfo {volume} {103}},\
  \bibinfo {pages} {1786} (\bibinfo {year} {1956})}\BibitemShut {NoStop}%
\bibitem [{\citenamefont {Belyaev}(1965)}]{Belyaev196517}%
  \BibitemOpen
  \bibfield  {author} {\bibinfo {author} {\bibfnamefont {S.~T.}\ \bibnamefont
  {Belyaev}},\ }\href {\doibase DOI: 10.1016/0029-5582(65)90840-0} {\bibfield
  {journal} {\bibinfo  {journal} {Nucl. Phys.}\ }\textbf {\bibinfo {volume}
  {64}},\ \bibinfo {pages} {17 } (\bibinfo {year} {1965})}\BibitemShut
  {NoStop}%
\bibitem [{\citenamefont {Libert}\ \emph {et~al.}(1999)\citenamefont {Libert},
  \citenamefont {Girod},\ and\ \citenamefont {Delaroche}}]{PhysRevC.60.054301}%
  \BibitemOpen
  \bibfield  {author} {\bibinfo {author} {\bibfnamefont {J.}~\bibnamefont
  {Libert}}, \bibinfo {author} {\bibfnamefont {M.}~\bibnamefont {Girod}}, \
  and\ \bibinfo {author} {\bibfnamefont {J.-P.}\ \bibnamefont {Delaroche}},\
  }\href {\doibase 10.1103/PhysRevC.60.054301} {\bibfield  {journal} {\bibinfo
  {journal} {Phys. Rev. C}\ }\textbf {\bibinfo {volume} {60}},\ \bibinfo
  {pages} {054301} (\bibinfo {year} {1999})}\BibitemShut {NoStop}%
\bibitem [{\citenamefont {Serra}\ \emph {et~al.}(2003)\citenamefont {Serra},
  \citenamefont {Nazmitdinov},\ and\ \citenamefont
  {Puente}}]{PhysRevB.68.035341}%
  \BibitemOpen
  \bibfield  {author} {\bibinfo {author} {\bibfnamefont {L.}~\bibnamefont
  {Serra}}, \bibinfo {author} {\bibfnamefont {R.~G.}\ \bibnamefont
  {Nazmitdinov}}, \ and\ \bibinfo {author} {\bibfnamefont {A.}~\bibnamefont
  {Puente}},\ }\href {\doibase 10.1103/PhysRevB.68.035341} {\bibfield
  {journal} {\bibinfo  {journal} {Phys. Rev. B}\ }\textbf {\bibinfo {volume}
  {68}},\ \bibinfo {pages} {035341} (\bibinfo {year} {2003})}\BibitemShut
  {NoStop}%
\bibitem [{\citenamefont {Bertsch}\ \emph {et~al.}(2007)\citenamefont
  {Bertsch}, \citenamefont {Girod}, \citenamefont {Hilaire}, \citenamefont
  {Delaroche}, \citenamefont {Goutte},\ and\ \citenamefont
  {P\'{e}ru}}]{bertsch:032502}%
  \BibitemOpen
  \bibfield  {author} {\bibinfo {author} {\bibfnamefont {G.~F.}\ \bibnamefont
  {Bertsch}}, \bibinfo {author} {\bibfnamefont {M.}~\bibnamefont {Girod}},
  \bibinfo {author} {\bibfnamefont {S.}~\bibnamefont {Hilaire}}, \bibinfo
  {author} {\bibfnamefont {J.-P.}\ \bibnamefont {Delaroche}}, \bibinfo {author}
  {\bibfnamefont {H.}~\bibnamefont {Goutte}}, \ and\ \bibinfo {author}
  {\bibfnamefont {S.}~\bibnamefont {P\'{e}ru}},\ }\href {\doibase
  10.1103/PhysRevLett.99.032502} {\bibfield  {journal} {\bibinfo  {journal}
  {Phys. Rev. Lett.}\ }\textbf {\bibinfo {volume} {99}},\ \bibinfo {eid}
  {032502} (\bibinfo {year} {2007})}\BibitemShut {NoStop}%
\bibitem [{\citenamefont {Yoshida}\ and\ \citenamefont
  {Yamagami}(2008)}]{yoshida:044312}%
  \BibitemOpen
  \bibfield  {author} {\bibinfo {author} {\bibfnamefont {K.}~\bibnamefont
  {Yoshida}}\ and\ \bibinfo {author} {\bibfnamefont {M.}~\bibnamefont
  {Yamagami}},\ }\href {\doibase 10.1103/PhysRevC.77.044312} {\bibfield
  {journal} {\bibinfo  {journal} {Phys. Rev. C}\ }\textbf {\bibinfo {volume}
  {77}},\ \bibinfo {eid} {044312} (\bibinfo {year} {2008})}\BibitemShut
  {NoStop}%
\bibitem [{\citenamefont {Yoshida}\ and\ \citenamefont
  {Giai}(2008)}]{yoshida:064316}%
  \BibitemOpen
  \bibfield  {author} {\bibinfo {author} {\bibfnamefont {K.}~\bibnamefont
  {Yoshida}}\ and\ \bibinfo {author} {\bibfnamefont {N.~V.}\ \bibnamefont
  {Giai}},\ }\href {\doibase 10.1103/PhysRevC.78.064316} {\bibfield  {journal}
  {\bibinfo  {journal} {Phys. Rev. C}\ }\textbf {\bibinfo {volume} {78}},\
  \bibinfo {eid} {064316} (\bibinfo {year} {2008})}\BibitemShut {NoStop}%
\bibitem [{\citenamefont {Yoshida}\ and\ \citenamefont
  {Hinohara}(2011)}]{PhysRevC.83.061302}%
  \BibitemOpen
  \bibfield  {author} {\bibinfo {author} {\bibfnamefont {K.}~\bibnamefont
  {Yoshida}}\ and\ \bibinfo {author} {\bibfnamefont {N.}~\bibnamefont
  {Hinohara}},\ }\href {\doibase 10.1103/PhysRevC.83.061302} {\bibfield
  {journal} {\bibinfo  {journal} {Phys. Rev. C}\ }\textbf {\bibinfo {volume}
  {83}},\ \bibinfo {pages} {061302} (\bibinfo {year} {2011})}\BibitemShut
  {NoStop}%
\bibitem [{\citenamefont {Afanasjev}\ \emph {et~al.}(2000)\citenamefont
  {Afanasjev}, \citenamefont {K\"onig}, \citenamefont {Ring}, \citenamefont
  {Robledo},\ and\ \citenamefont {Egido}}]{PhysRevC.62.054306}%
  \BibitemOpen
  \bibfield  {author} {\bibinfo {author} {\bibfnamefont {A.~V.}\ \bibnamefont
  {Afanasjev}}, \bibinfo {author} {\bibfnamefont {J.}~\bibnamefont {K\"onig}},
  \bibinfo {author} {\bibfnamefont {P.}~\bibnamefont {Ring}}, \bibinfo {author}
  {\bibfnamefont {L.~M.}\ \bibnamefont {Robledo}}, \ and\ \bibinfo {author}
  {\bibfnamefont {J.~L.}\ \bibnamefont {Egido}},\ }\href {\doibase
  10.1103/PhysRevC.62.054306} {\bibfield  {journal} {\bibinfo  {journal} {Phys.
  Rev. C}\ }\textbf {\bibinfo {volume} {62}},\ \bibinfo {pages} {054306}
  (\bibinfo {year} {2000})}\BibitemShut {NoStop}%
\bibitem [{\citenamefont {Delaroche}\ \emph {et~al.}(2010)\citenamefont
  {Delaroche}, \citenamefont {Girod}, \citenamefont {Libert}, \citenamefont
  {Goutte}, \citenamefont {Hilaire}, \citenamefont {P\'eru}, \citenamefont
  {Pillet},\ and\ \citenamefont {Bertsch}}]{PhysRevC.81.014303}%
  \BibitemOpen
  \bibfield  {author} {\bibinfo {author} {\bibfnamefont {J.~P.}\ \bibnamefont
  {Delaroche}}, \bibinfo {author} {\bibfnamefont {M.}~\bibnamefont {Girod}},
  \bibinfo {author} {\bibfnamefont {J.}~\bibnamefont {Libert}}, \bibinfo
  {author} {\bibfnamefont {H.}~\bibnamefont {Goutte}}, \bibinfo {author}
  {\bibfnamefont {S.}~\bibnamefont {Hilaire}}, \bibinfo {author} {\bibfnamefont
  {S.}~\bibnamefont {P\'eru}}, \bibinfo {author} {\bibfnamefont
  {N.}~\bibnamefont {Pillet}}, \ and\ \bibinfo {author} {\bibfnamefont {G.~F.}\
  \bibnamefont {Bertsch}},\ }\href {\doibase 10.1103/PhysRevC.81.014303}
  {\bibfield  {journal} {\bibinfo  {journal} {Phys. Rev. C}\ }\textbf {\bibinfo
  {volume} {81}},\ \bibinfo {pages} {014303} (\bibinfo {year}
  {2010})}\BibitemShut {NoStop}%
\bibitem [{\citenamefont {Li}\ \emph {et~al.}(2012)\citenamefont {Li},
  \citenamefont {Nik\ifmmode \check{s}\else \v{s}\fi{}i\ifmmode~\acute{c}\else
  \'{c}\fi{}}, \citenamefont {Ring}, \citenamefont {Vretenar}, \citenamefont
  {Yao},\ and\ \citenamefont {Meng}}]{PhysRevC.86.034334}%
  \BibitemOpen
  \bibfield  {author} {\bibinfo {author} {\bibfnamefont {Z.~P.}\ \bibnamefont
  {Li}}, \bibinfo {author} {\bibfnamefont {T.}~\bibnamefont {Nik\ifmmode
  \check{s}\else \v{s}\fi{}i\ifmmode~\acute{c}\else \'{c}\fi{}}}, \bibinfo
  {author} {\bibfnamefont {P.}~\bibnamefont {Ring}}, \bibinfo {author}
  {\bibfnamefont {D.}~\bibnamefont {Vretenar}}, \bibinfo {author}
  {\bibfnamefont {J.~M.}\ \bibnamefont {Yao}}, \ and\ \bibinfo {author}
  {\bibfnamefont {J.}~\bibnamefont {Meng}},\ }\href {\doibase
  10.1103/PhysRevC.86.034334} {\bibfield  {journal} {\bibinfo  {journal} {Phys.
  Rev. C}\ }\textbf {\bibinfo {volume} {86}},\ \bibinfo {pages} {034334}
  (\bibinfo {year} {2012})}\BibitemShut {NoStop}%
\bibitem [{\citenamefont {Nakatsukasa}\ \emph {et~al.}(2007)\citenamefont
  {Nakatsukasa}, \citenamefont {Inakura},\ and\ \citenamefont
  {Yabana}}]{nakatsukasa:024318}%
  \BibitemOpen
  \bibfield  {author} {\bibinfo {author} {\bibfnamefont {T.}~\bibnamefont
  {Nakatsukasa}}, \bibinfo {author} {\bibfnamefont {T.}~\bibnamefont
  {Inakura}}, \ and\ \bibinfo {author} {\bibfnamefont {K.}~\bibnamefont
  {Yabana}},\ }\href {\doibase 10.1103/PhysRevC.76.024318} {\bibfield
  {journal} {\bibinfo  {journal} {Phys. Rev. C}\ }\textbf {\bibinfo {volume}
  {76}},\ \bibinfo {eid} {024318} (\bibinfo {year} {2007})}\BibitemShut
  {NoStop}%
\bibitem [{\citenamefont {Stoitsov}\ \emph {et~al.}(2011)\citenamefont
  {Stoitsov}, \citenamefont {Kortelainen}, \citenamefont {Nakatsukasa},
  \citenamefont {Losa},\ and\ \citenamefont {Nazarewicz}}]{PhysRevC.84.041305}%
  \BibitemOpen
  \bibfield  {author} {\bibinfo {author} {\bibfnamefont {M.}~\bibnamefont
  {Stoitsov}}, \bibinfo {author} {\bibfnamefont {M.}~\bibnamefont
  {Kortelainen}}, \bibinfo {author} {\bibfnamefont {T.}~\bibnamefont
  {Nakatsukasa}}, \bibinfo {author} {\bibfnamefont {C.}~\bibnamefont {Losa}}, \
  and\ \bibinfo {author} {\bibfnamefont {W.}~\bibnamefont {Nazarewicz}},\
  }\href {\doibase 10.1103/PhysRevC.84.041305} {\bibfield  {journal} {\bibinfo
  {journal} {Phys. Rev. C}\ }\textbf {\bibinfo {volume} {84}},\ \bibinfo
  {pages} {041305} (\bibinfo {year} {2011})}\BibitemShut {NoStop}%
\bibitem [{\citenamefont {Avogadro}\ and\ \citenamefont
  {Nakatsukasa}(2011)}]{PhysRevC.84.014314}%
  \BibitemOpen
  \bibfield  {author} {\bibinfo {author} {\bibfnamefont {P.}~\bibnamefont
  {Avogadro}}\ and\ \bibinfo {author} {\bibfnamefont {T.}~\bibnamefont
  {Nakatsukasa}},\ }\href {\doibase 10.1103/PhysRevC.84.014314} {\bibfield
  {journal} {\bibinfo  {journal} {Phys. Rev. C}\ }\textbf {\bibinfo {volume}
  {84}},\ \bibinfo {pages} {014314} (\bibinfo {year} {2011})}\BibitemShut
  {NoStop}%
\bibitem [{\citenamefont {Pei}\ \emph {et~al.}(2014)\citenamefont {Pei},
  \citenamefont {Kortelainen}, \citenamefont {Zhang},\ and\ \citenamefont
  {Xu}}]{PhysRevC.90.051304}%
  \BibitemOpen
  \bibfield  {author} {\bibinfo {author} {\bibfnamefont {J.~C.}\ \bibnamefont
  {Pei}}, \bibinfo {author} {\bibfnamefont {M.}~\bibnamefont {Kortelainen}},
  \bibinfo {author} {\bibfnamefont {Y.~N.}\ \bibnamefont {Zhang}}, \ and\
  \bibinfo {author} {\bibfnamefont {F.~R.}\ \bibnamefont {Xu}},\ }\href
  {\doibase 10.1103/PhysRevC.90.051304} {\bibfield  {journal} {\bibinfo
  {journal} {Phys. Rev. C}\ }\textbf {\bibinfo {volume} {90}},\ \bibinfo
  {pages} {051304} (\bibinfo {year} {2014})}\BibitemShut {NoStop}%
\bibitem [{\citenamefont {Liang}\ \emph {et~al.}(2013)\citenamefont {Liang},
  \citenamefont {Nakatsukasa}, \citenamefont {Niu},\ and\ \citenamefont
  {Meng}}]{PhysRevC.87.054310}%
  \BibitemOpen
  \bibfield  {author} {\bibinfo {author} {\bibfnamefont {H.}~\bibnamefont
  {Liang}}, \bibinfo {author} {\bibfnamefont {T.}~\bibnamefont {Nakatsukasa}},
  \bibinfo {author} {\bibfnamefont {Z.}~\bibnamefont {Niu}}, \ and\ \bibinfo
  {author} {\bibfnamefont {J.}~\bibnamefont {Meng}},\ }\href {\doibase
  10.1103/PhysRevC.87.054310} {\bibfield  {journal} {\bibinfo  {journal} {Phys.
  Rev. C}\ }\textbf {\bibinfo {volume} {87}},\ \bibinfo {pages} {054310}
  (\bibinfo {year} {2013})}\BibitemShut {NoStop}%
\bibitem [{\citenamefont {Nik\ifmmode \check{s}\else
  \v{s}\fi{}i\ifmmode~\acute{c}\else \'{c}\fi{}}\ \emph
  {et~al.}(2013)\citenamefont {Nik\ifmmode \check{s}\else
  \v{s}\fi{}i\ifmmode~\acute{c}\else \'{c}\fi{}}, \citenamefont {Kralj},
  \citenamefont {Tuti\ifmmode~\check{s}\else \v{s}\fi{}}, \citenamefont
  {Vretenar},\ and\ \citenamefont {Ring}}]{PhysRevC.88.044327}%
  \BibitemOpen
  \bibfield  {author} {\bibinfo {author} {\bibfnamefont {T.}~\bibnamefont
  {Nik\ifmmode \check{s}\else \v{s}\fi{}i\ifmmode~\acute{c}\else \'{c}\fi{}}},
  \bibinfo {author} {\bibfnamefont {N.}~\bibnamefont {Kralj}}, \bibinfo
  {author} {\bibfnamefont {T.}~\bibnamefont {Tuti\ifmmode~\check{s}\else
  \v{s}\fi{}}}, \bibinfo {author} {\bibfnamefont {D.}~\bibnamefont {Vretenar}},
  \ and\ \bibinfo {author} {\bibfnamefont {P.}~\bibnamefont {Ring}},\ }\href
  {\doibase 10.1103/PhysRevC.88.044327} {\bibfield  {journal} {\bibinfo
  {journal} {Phys. Rev. C}\ }\textbf {\bibinfo {volume} {88}},\ \bibinfo
  {pages} {044327} (\bibinfo {year} {2013})}\BibitemShut {NoStop}%
\bibitem [{\citenamefont {Inakura}\ \emph {et~al.}(2011)\citenamefont
  {Inakura}, \citenamefont {Nakatsukasa},\ and\ \citenamefont
  {Yabana}}]{PhysRevC.84.021302}%
  \BibitemOpen
  \bibfield  {author} {\bibinfo {author} {\bibfnamefont {T.}~\bibnamefont
  {Inakura}}, \bibinfo {author} {\bibfnamefont {T.}~\bibnamefont
  {Nakatsukasa}}, \ and\ \bibinfo {author} {\bibfnamefont {K.}~\bibnamefont
  {Yabana}},\ }\href {\doibase 10.1103/PhysRevC.84.021302} {\bibfield
  {journal} {\bibinfo  {journal} {Phys. Rev. C}\ }\textbf {\bibinfo {volume}
  {84}},\ \bibinfo {pages} {021302} (\bibinfo {year} {2011})}\BibitemShut
  {NoStop}%
\bibitem [{\citenamefont {Inakura}\ \emph {et~al.}(2014)\citenamefont
  {Inakura}, \citenamefont {Horiuchi}, \citenamefont {Suzuki},\ and\
  \citenamefont {Nakatsukasa}}]{PhysRevC.89.064316}%
  \BibitemOpen
  \bibfield  {author} {\bibinfo {author} {\bibfnamefont {T.}~\bibnamefont
  {Inakura}}, \bibinfo {author} {\bibfnamefont {W.}~\bibnamefont {Horiuchi}},
  \bibinfo {author} {\bibfnamefont {Y.}~\bibnamefont {Suzuki}}, \ and\ \bibinfo
  {author} {\bibfnamefont {T.}~\bibnamefont {Nakatsukasa}},\ }\href {\doibase
  10.1103/PhysRevC.89.064316} {\bibfield  {journal} {\bibinfo  {journal} {Phys.
  Rev. C}\ }\textbf {\bibinfo {volume} {89}},\ \bibinfo {pages} {064316}
  (\bibinfo {year} {2014})}\BibitemShut {NoStop}%
\bibitem [{\citenamefont {Avogadro}\ and\ \citenamefont
  {Nakatsukasa}(2013)}]{PhysRevC.87.014331}%
  \BibitemOpen
  \bibfield  {author} {\bibinfo {author} {\bibfnamefont {P.}~\bibnamefont
  {Avogadro}}\ and\ \bibinfo {author} {\bibfnamefont {T.}~\bibnamefont
  {Nakatsukasa}},\ }\href {\doibase 10.1103/PhysRevC.87.014331} {\bibfield
  {journal} {\bibinfo  {journal} {Phys. Rev. C}\ }\textbf {\bibinfo {volume}
  {87}},\ \bibinfo {pages} {014331} (\bibinfo {year} {2013})}\BibitemShut
  {NoStop}%
\bibitem [{\citenamefont {Hinohara}\ \emph {et~al.}(2013)\citenamefont
  {Hinohara}, \citenamefont {Kortelainen},\ and\ \citenamefont
  {Nazarewicz}}]{PhysRevC.87.064309}%
  \BibitemOpen
  \bibfield  {author} {\bibinfo {author} {\bibfnamefont {N.}~\bibnamefont
  {Hinohara}}, \bibinfo {author} {\bibfnamefont {M.}~\bibnamefont
  {Kortelainen}}, \ and\ \bibinfo {author} {\bibfnamefont {W.}~\bibnamefont
  {Nazarewicz}},\ }\href {\doibase 10.1103/PhysRevC.87.064309} {\bibfield
  {journal} {\bibinfo  {journal} {Phys. Rev. C}\ }\textbf {\bibinfo {volume}
  {87}},\ \bibinfo {pages} {064309} (\bibinfo {year} {2013})}\BibitemShut
  {NoStop}%
\bibitem [{\citenamefont {Hinohara}\ \emph {et~al.}(2015)\citenamefont
  {Hinohara}, \citenamefont {Kortelainen}, \citenamefont {Nazarewicz},\ and\
  \citenamefont {Olsen}}]{PhysRevC.91.044323}%
  \BibitemOpen
  \bibfield  {author} {\bibinfo {author} {\bibfnamefont {N.}~\bibnamefont
  {Hinohara}}, \bibinfo {author} {\bibfnamefont {M.}~\bibnamefont
  {Kortelainen}}, \bibinfo {author} {\bibfnamefont {W.}~\bibnamefont
  {Nazarewicz}}, \ and\ \bibinfo {author} {\bibfnamefont {E.}~\bibnamefont
  {Olsen}},\ }\href {\doibase 10.1103/PhysRevC.91.044323} {\bibfield  {journal}
  {\bibinfo  {journal} {Phys. Rev. C}\ }\textbf {\bibinfo {volume} {91}},\
  \bibinfo {pages} {044323} (\bibinfo {year} {2015})}\BibitemShut {NoStop}%
\bibitem [{\citenamefont {Toivanen}\ \emph {et~al.}(2010)\citenamefont
  {Toivanen}, \citenamefont {Carlsson}, \citenamefont {Dobaczewski},
  \citenamefont {Mizuyama}, \citenamefont {Rodr\'\i{}guez-Guzm\'an},
  \citenamefont {Toivanen},\ and\ \citenamefont
  {Vesel\'y}}]{PhysRevC.81.034312}%
  \BibitemOpen
  \bibfield  {author} {\bibinfo {author} {\bibfnamefont {J.}~\bibnamefont
  {Toivanen}}, \bibinfo {author} {\bibfnamefont {B.~G.}\ \bibnamefont
  {Carlsson}}, \bibinfo {author} {\bibfnamefont {J.}~\bibnamefont
  {Dobaczewski}}, \bibinfo {author} {\bibfnamefont {K.}~\bibnamefont
  {Mizuyama}}, \bibinfo {author} {\bibfnamefont {R.~R.}\ \bibnamefont
  {Rodr\'\i{}guez-Guzm\'an}}, \bibinfo {author} {\bibfnamefont
  {P.}~\bibnamefont {Toivanen}}, \ and\ \bibinfo {author} {\bibfnamefont
  {P.}~\bibnamefont {Vesel\'y}},\ }\href {\doibase 10.1103/PhysRevC.81.034312}
  {\bibfield  {journal} {\bibinfo  {journal} {Phys. Rev. C}\ }\textbf {\bibinfo
  {volume} {81}},\ \bibinfo {pages} {034312} (\bibinfo {year}
  {2010})}\BibitemShut {NoStop}%
\bibitem [{\citenamefont {Marshalek}\ and\ \citenamefont
  {Weneser}(1969)}]{Marshalek1969}%
  \BibitemOpen
  \bibfield  {author} {\bibinfo {author} {\bibfnamefont {E.~R.}\ \bibnamefont
  {Marshalek}}\ and\ \bibinfo {author} {\bibfnamefont {J.}~\bibnamefont
  {Weneser}},\ }\href {\doibase http://dx.doi.org/10.1016/0003-4916(69)90037-2}
  {\bibfield  {journal} {\bibinfo  {journal} {Ann. Phys.}\ }\textbf {\bibinfo
  {volume} {53}},\ \bibinfo {pages} {569 } (\bibinfo {year}
  {1969})}\BibitemShut {NoStop}%
\bibitem [{\citenamefont {Yoshida}(2014)}]{PhysRevC.90.031303}%
  \BibitemOpen
  \bibfield  {author} {\bibinfo {author} {\bibfnamefont {K.}~\bibnamefont
  {Yoshida}},\ }\href {\doibase 10.1103/PhysRevC.90.031303} {\bibfield
  {journal} {\bibinfo  {journal} {Phys. Rev. C}\ }\textbf {\bibinfo {volume}
  {90}},\ \bibinfo {pages} {031303} (\bibinfo {year} {2014})}\BibitemShut
  {NoStop}%
\bibitem [{\citenamefont {Matsuo}\ \emph {et~al.}(2000)\citenamefont {Matsuo},
  \citenamefont {Nakatsukasa},\ and\ \citenamefont
  {Matsuyanagi}}]{PTP.103.959}%
  \BibitemOpen
  \bibfield  {author} {\bibinfo {author} {\bibfnamefont {M.}~\bibnamefont
  {Matsuo}}, \bibinfo {author} {\bibfnamefont {T.}~\bibnamefont {Nakatsukasa}},
  \ and\ \bibinfo {author} {\bibfnamefont {K.}~\bibnamefont {Matsuyanagi}},\
  }\href {\doibase 10.1143/PTP.103.959} {\bibfield  {journal} {\bibinfo
  {journal} {Prog. Theor. Phys.}\ }\textbf {\bibinfo {volume} {103}},\ \bibinfo
  {pages} {959} (\bibinfo {year} {2000})}\BibitemShut {NoStop}%
\bibitem [{\citenamefont {Blaizot}\ and\ \citenamefont
  {Ripka}(1986)}]{Blaizot-Ripka}%
  \BibitemOpen
  \bibfield  {author} {\bibinfo {author} {\bibfnamefont {J.-P.}\ \bibnamefont
  {Blaizot}}\ and\ \bibinfo {author} {\bibfnamefont {G.}~\bibnamefont
  {Ripka}},\ }\href@noop {} {\emph {\bibinfo {title} {Quantum Theory of Finite
  Systems}}}\ (\bibinfo  {publisher} {The MIT press},\ \bibinfo {year}
  {1986})\BibitemShut {NoStop}%
\bibitem [{\citenamefont {Capelli}\ \emph {et~al.}(2009)\citenamefont
  {Capelli}, \citenamefont {Col\`o},\ and\ \citenamefont
  {Li}}]{PhysRevC.79.054329}%
  \BibitemOpen
  \bibfield  {author} {\bibinfo {author} {\bibfnamefont {L.}~\bibnamefont
  {Capelli}}, \bibinfo {author} {\bibfnamefont {G.}~\bibnamefont {Col\`o}}, \
  and\ \bibinfo {author} {\bibfnamefont {J.}~\bibnamefont {Li}},\ }\href
  {\doibase 10.1103/PhysRevC.79.054329} {\bibfield  {journal} {\bibinfo
  {journal} {Phys. Rev. C}\ }\textbf {\bibinfo {volume} {79}},\ \bibinfo
  {pages} {054329} (\bibinfo {year} {2009})}\BibitemShut {NoStop}%
\bibitem [{\citenamefont {Nazmitdinov}\ \emph {et~al.}(2002)\citenamefont
  {Nazmitdinov}, \citenamefont {Almehed},\ and\ \citenamefont
  {D\"onau}}]{PhysRevC.65.041307}%
  \BibitemOpen
  \bibfield  {author} {\bibinfo {author} {\bibfnamefont {R.~G.}\ \bibnamefont
  {Nazmitdinov}}, \bibinfo {author} {\bibfnamefont {D.}~\bibnamefont
  {Almehed}}, \ and\ \bibinfo {author} {\bibfnamefont {F.}~\bibnamefont
  {D\"onau}},\ }\href {\doibase 10.1103/PhysRevC.65.041307} {\bibfield
  {journal} {\bibinfo  {journal} {Phys. Rev. C}\ }\textbf {\bibinfo {volume}
  {65}},\ \bibinfo {pages} {041307} (\bibinfo {year} {2002})}\BibitemShut
  {NoStop}%
\bibitem [{\citenamefont {Stetcu}\ and\ \citenamefont
  {Johnson}(2003)}]{PhysRevC.67.044315}%
  \BibitemOpen
  \bibfield  {author} {\bibinfo {author} {\bibfnamefont {I.}~\bibnamefont
  {Stetcu}}\ and\ \bibinfo {author} {\bibfnamefont {C.~W.}\ \bibnamefont
  {Johnson}},\ }\href {\doibase 10.1103/PhysRevC.67.044315} {\bibfield
  {journal} {\bibinfo  {journal} {Phys. Rev. C}\ }\textbf {\bibinfo {volume}
  {67}},\ \bibinfo {pages} {044315} (\bibinfo {year} {2003})}\BibitemShut
  {NoStop}%
\bibitem [{\citenamefont {Chabanat}\ \emph {et~al.}(1998)\citenamefont
  {Chabanat}, \citenamefont {Bonche}, \citenamefont {Haensel}, \citenamefont
  {Meyer},\ and\ \citenamefont {Schaeffer}}]{Chabanat1998231}%
  \BibitemOpen
  \bibfield  {author} {\bibinfo {author} {\bibfnamefont {E.}~\bibnamefont
  {Chabanat}}, \bibinfo {author} {\bibfnamefont {P.}~\bibnamefont {Bonche}},
  \bibinfo {author} {\bibfnamefont {P.}~\bibnamefont {Haensel}}, \bibinfo
  {author} {\bibfnamefont {J.}~\bibnamefont {Meyer}}, \ and\ \bibinfo {author}
  {\bibfnamefont {R.}~\bibnamefont {Schaeffer}},\ }\href {\doibase
  10.1016/S0375-9474(98)00180-8} {\bibfield  {journal} {\bibinfo  {journal}
  {Nucl. Phys. A}\ }\textbf {\bibinfo {volume} {635}},\ \bibinfo {pages} {231 }
  (\bibinfo {year} {1998})}\BibitemShut {NoStop}%
\bibitem [{\citenamefont {Stoitsov}\ \emph {et~al.}(2005)\citenamefont
  {Stoitsov}, \citenamefont {Dobaczewski}, \citenamefont {Nazarewicz},\ and\
  \citenamefont {Ring}}]{Stoitsov200543}%
  \BibitemOpen
  \bibfield  {author} {\bibinfo {author} {\bibfnamefont {M.}~\bibnamefont
  {Stoitsov}}, \bibinfo {author} {\bibfnamefont {J.}~\bibnamefont
  {Dobaczewski}}, \bibinfo {author} {\bibfnamefont {W.}~\bibnamefont
  {Nazarewicz}}, \ and\ \bibinfo {author} {\bibfnamefont {P.}~\bibnamefont
  {Ring}},\ }\href {\doibase DOI: 10.1016/j.cpc.2005.01.001} {\bibfield
  {journal} {\bibinfo  {journal} {Comp. Phys. Comm.}\ }\textbf {\bibinfo
  {volume} {167}},\ \bibinfo {pages} {43 } (\bibinfo {year}
  {2005})}\BibitemShut {NoStop}%
\bibitem [{\citenamefont {Stoitsov}\ \emph {et~al.}(2013)\citenamefont
  {Stoitsov}, \citenamefont {Schunck}, \citenamefont {Kortelainen},
  \citenamefont {Michel}, \citenamefont {Nam}, \citenamefont {Olsen},
  \citenamefont {Sarich},\ and\ \citenamefont {Wild}}]{Stoitsov20131592}%
  \BibitemOpen
  \bibfield  {author} {\bibinfo {author} {\bibfnamefont {M.}~\bibnamefont
  {Stoitsov}}, \bibinfo {author} {\bibfnamefont {N.}~\bibnamefont {Schunck}},
  \bibinfo {author} {\bibfnamefont {M.}~\bibnamefont {Kortelainen}}, \bibinfo
  {author} {\bibfnamefont {N.}~\bibnamefont {Michel}}, \bibinfo {author}
  {\bibfnamefont {H.}~\bibnamefont {Nam}}, \bibinfo {author} {\bibfnamefont
  {E.}~\bibnamefont {Olsen}}, \bibinfo {author} {\bibfnamefont
  {J.}~\bibnamefont {Sarich}}, \ and\ \bibinfo {author} {\bibfnamefont
  {S.}~\bibnamefont {Wild}},\ }\href {\doibase 10.1016/j.cpc.2013.01.013}
  {\bibfield  {journal} {\bibinfo  {journal} {Comp. Phys. Comm.}\ }\textbf
  {\bibinfo {volume} {184}},\ \bibinfo {pages} {1592} (\bibinfo {year}
  {2013})}\BibitemShut {NoStop}%
\bibitem [{\citenamefont {Beiner}\ \emph {et~al.}(1975)\citenamefont {Beiner},
  \citenamefont {Flocard}, \citenamefont {Giai},\ and\ \citenamefont
  {Quentin}}]{Beiner197529}%
  \BibitemOpen
  \bibfield  {author} {\bibinfo {author} {\bibfnamefont {M.}~\bibnamefont
  {Beiner}}, \bibinfo {author} {\bibfnamefont {H.}~\bibnamefont {Flocard}},
  \bibinfo {author} {\bibfnamefont {N.~V.}\ \bibnamefont {Giai}}, \ and\
  \bibinfo {author} {\bibfnamefont {P.}~\bibnamefont {Quentin}},\ }\href
  {\doibase http://dx.doi.org/10.1016/0375-9474(75)90338-3} {\bibfield
  {journal} {\bibinfo  {journal} {Nucl. Phys. A}\ }\textbf {\bibinfo {volume}
  {238}},\ \bibinfo {pages} {29 } (\bibinfo {year} {1975})}\BibitemShut
  {NoStop}%
\bibitem [{\citenamefont {Bender}\ \emph {et~al.}(2000)\citenamefont {Bender},
  \citenamefont {Rutz}, \citenamefont {Reinhard},\ and\ \citenamefont
  {Maruhn}}]{EPJA7_467}%
  \BibitemOpen
  \bibfield  {author} {\bibinfo {author} {\bibfnamefont {M.}~\bibnamefont
  {Bender}}, \bibinfo {author} {\bibfnamefont {K.}~\bibnamefont {Rutz}},
  \bibinfo {author} {\bibfnamefont {P.-G.}\ \bibnamefont {Reinhard}}, \ and\
  \bibinfo {author} {\bibfnamefont {J.}~\bibnamefont {Maruhn}},\ }\href
  {\doibase 10.1007/PL00013645} {\bibfield  {journal} {\bibinfo  {journal}
  {Eur. Phys. J. A}\ }\textbf {\bibinfo {volume} {7}},\ \bibinfo {pages} {467}
  (\bibinfo {year} {2000})}\BibitemShut {NoStop}%
\bibitem [{\citenamefont {Dobaczewski}(2009)}]{0954-3899-36-10-105105}%
  \BibitemOpen
  \bibfield  {author} {\bibinfo {author} {\bibfnamefont {J.}~\bibnamefont
  {Dobaczewski}},\ }\href {http://stacks.iop.org/0954-3899/36/i=10/a=105105}
  {\bibfield  {journal} {\bibinfo  {journal} {J. Phys. G}\ }\textbf {\bibinfo
  {volume} {36}},\ \bibinfo {pages} {105105} (\bibinfo {year}
  {2009})}\BibitemShut {NoStop}%
\bibitem [{\citenamefont {Broglia}\ \emph {et~al.}(2000)\citenamefont
  {Broglia}, \citenamefont {Terasaki},\ and\ \citenamefont
  {Giovanardi}}]{Broglia20001}%
  \BibitemOpen
  \bibfield  {author} {\bibinfo {author} {\bibfnamefont {R.}~\bibnamefont
  {Broglia}}, \bibinfo {author} {\bibfnamefont {J.}~\bibnamefont {Terasaki}}, \
  and\ \bibinfo {author} {\bibfnamefont {N.}~\bibnamefont {Giovanardi}},\
  }\href {\doibase http://dx.doi.org/10.1016/S0370-1573(00)00046-6} {\bibfield
  {journal} {\bibinfo  {journal} {Phys. Rep.}\ }\textbf {\bibinfo {volume}
  {335}},\ \bibinfo {pages} {1 } (\bibinfo {year} {2000})}\BibitemShut
  {NoStop}%
\bibitem [{\citenamefont {Brink}\ and\ \citenamefont
  {Broglia}(2005)}]{Brink-Broglia}%
  \BibitemOpen
  \bibfield  {author} {\bibinfo {author} {\bibfnamefont {D.~M.}\ \bibnamefont
  {Brink}}\ and\ \bibinfo {author} {\bibfnamefont {R.~A.}\ \bibnamefont
  {Broglia}},\ }\href@noop {} {\emph {\bibinfo {title} {Nuclear Superfluidity,
  Pairing in Finite Systems}}}\ (\bibinfo  {publisher} {Cambridge University
  Press},\ \bibinfo {year} {2005})\BibitemShut {NoStop}%
\bibitem [{\citenamefont {Shimoyama}\ and\ \citenamefont
  {Matsuo}(2011)}]{PhysRevC.84.044317}%
  \BibitemOpen
  \bibfield  {author} {\bibinfo {author} {\bibfnamefont {H.}~\bibnamefont
  {Shimoyama}}\ and\ \bibinfo {author} {\bibfnamefont {M.}~\bibnamefont
  {Matsuo}},\ }\href {\doibase 10.1103/PhysRevC.84.044317} {\bibfield
  {journal} {\bibinfo  {journal} {Phys. Rev. C}\ }\textbf {\bibinfo {volume}
  {84}},\ \bibinfo {pages} {044317} (\bibinfo {year} {2011})}\BibitemShut
  {NoStop}%
\bibitem [{\citenamefont {Shimoyama}\ and\ \citenamefont
  {Matsuo}(2013)}]{PhysRevC.88.054308}%
  \BibitemOpen
  \bibfield  {author} {\bibinfo {author} {\bibfnamefont {H.}~\bibnamefont
  {Shimoyama}}\ and\ \bibinfo {author} {\bibfnamefont {M.}~\bibnamefont
  {Matsuo}},\ }\href {\doibase 10.1103/PhysRevC.88.054308} {\bibfield
  {journal} {\bibinfo  {journal} {Phys. Rev. C}\ }\textbf {\bibinfo {volume}
  {88}},\ \bibinfo {pages} {054308} (\bibinfo {year} {2013})}\BibitemShut
  {NoStop}%
\bibitem [{\citenamefont {Potel}\ \emph {et~al.}(2011)\citenamefont {Potel},
  \citenamefont {Barranco}, \citenamefont {Marini}, \citenamefont {Idini},
  \citenamefont {Vigezzi},\ and\ \citenamefont
  {Broglia}}]{PhysRevLett.107.092501}%
  \BibitemOpen
  \bibfield  {author} {\bibinfo {author} {\bibfnamefont {G.}~\bibnamefont
  {Potel}}, \bibinfo {author} {\bibfnamefont {F.}~\bibnamefont {Barranco}},
  \bibinfo {author} {\bibfnamefont {F.}~\bibnamefont {Marini}}, \bibinfo
  {author} {\bibfnamefont {A.}~\bibnamefont {Idini}}, \bibinfo {author}
  {\bibfnamefont {E.}~\bibnamefont {Vigezzi}}, \ and\ \bibinfo {author}
  {\bibfnamefont {R.~A.}\ \bibnamefont {Broglia}},\ }\href {\doibase
  10.1103/PhysRevLett.107.092501} {\bibfield  {journal} {\bibinfo  {journal}
  {Phys. Rev. Lett.}\ }\textbf {\bibinfo {volume} {107}},\ \bibinfo {pages}
  {092501} (\bibinfo {year} {2011})},\ \bibinfo {note} {{\bf 108}, 069904
  (2012)}\BibitemShut {NoStop}%
\bibitem [{\citenamefont {Potel}\ \emph {et~al.}(2013)\citenamefont {Potel},
  \citenamefont {Idini}, \citenamefont {Barranco}, \citenamefont {Vigezzi},\
  and\ \citenamefont {Broglia}}]{PhysRevC.87.054321}%
  \BibitemOpen
  \bibfield  {author} {\bibinfo {author} {\bibfnamefont {G.}~\bibnamefont
  {Potel}}, \bibinfo {author} {\bibfnamefont {A.}~\bibnamefont {Idini}},
  \bibinfo {author} {\bibfnamefont {F.}~\bibnamefont {Barranco}}, \bibinfo
  {author} {\bibfnamefont {E.}~\bibnamefont {Vigezzi}}, \ and\ \bibinfo
  {author} {\bibfnamefont {R.~A.}\ \bibnamefont {Broglia}},\ }\href {\doibase
  10.1103/PhysRevC.87.054321} {\bibfield  {journal} {\bibinfo  {journal} {Phys.
  Rev. C}\ }\textbf {\bibinfo {volume} {87}},\ \bibinfo {pages} {054321}
  (\bibinfo {year} {2013})}\BibitemShut {NoStop}%
\bibitem [{\citenamefont {Audi}\ \emph {et~al.}(2012)\citenamefont {Audi},
  \citenamefont {Wang}, \citenamefont {Wapstra}, \citenamefont {Kondev},
  \citenamefont {MacCormick}, \citenamefont {Xu},\ and\ \citenamefont
  {Pfeiffer}}]{ame2012}%
  \BibitemOpen
  \bibfield  {author} {\bibinfo {author} {\bibfnamefont {G.}~\bibnamefont
  {Audi}}, \bibinfo {author} {\bibfnamefont {M.}~\bibnamefont {Wang}}, \bibinfo
  {author} {\bibfnamefont {A.}~\bibnamefont {Wapstra}}, \bibinfo {author}
  {\bibfnamefont {F.}~\bibnamefont {Kondev}}, \bibinfo {author} {\bibfnamefont
  {M.}~\bibnamefont {MacCormick}}, \bibinfo {author} {\bibfnamefont
  {X.}~\bibnamefont {Xu}}, \ and\ \bibinfo {author} {\bibfnamefont
  {B.}~\bibnamefont {Pfeiffer}},\ }\href
  {http://stacks.iop.org/1674-1137/36/i=12/a=002} {\bibfield  {journal}
  {\bibinfo  {journal} {Chin. Phys. C}\ }\textbf {\bibinfo {volume} {36}},\
  \bibinfo {pages} {1287} (\bibinfo {year} {2012})}\BibitemShut {NoStop}%
\bibitem [{\citenamefont {Zalewski}\ \emph {et~al.}(2008)\citenamefont
  {Zalewski}, \citenamefont {Dobaczewski}, \citenamefont {Satu\l{}a},\ and\
  \citenamefont {Werner}}]{PhysRevC.77.024316}%
  \BibitemOpen
  \bibfield  {author} {\bibinfo {author} {\bibfnamefont {M.}~\bibnamefont
  {Zalewski}}, \bibinfo {author} {\bibfnamefont {J.}~\bibnamefont
  {Dobaczewski}}, \bibinfo {author} {\bibfnamefont {W.}~\bibnamefont
  {Satu\l{}a}}, \ and\ \bibinfo {author} {\bibfnamefont {T.~R.}\ \bibnamefont
  {Werner}},\ }\href {\doibase 10.1103/PhysRevC.77.024316} {\bibfield
  {journal} {\bibinfo  {journal} {Phys. Rev. C}\ }\textbf {\bibinfo {volume}
  {77}},\ \bibinfo {pages} {024316} (\bibinfo {year} {2008})}\BibitemShut
  {NoStop}%
\bibitem [{\citenamefont {Anguiano}\ \emph {et~al.}(2001)\citenamefont
  {Anguiano}, \citenamefont {Egido},\ and\ \citenamefont
  {Robledo}}]{Anguiano2001227}%
  \BibitemOpen
  \bibfield  {author} {\bibinfo {author} {\bibfnamefont {M.}~\bibnamefont
  {Anguiano}}, \bibinfo {author} {\bibfnamefont {J.}~\bibnamefont {Egido}}, \
  and\ \bibinfo {author} {\bibfnamefont {L.}~\bibnamefont {Robledo}},\ }\href
  {\doibase http://dx.doi.org/10.1016/S0375-9474(00)00445-0} {\bibfield
  {journal} {\bibinfo  {journal} {Nucl. Phys. A}\ }\textbf {\bibinfo {volume}
  {683}},\ \bibinfo {pages} {227 } (\bibinfo {year} {2001})}\BibitemShut
  {NoStop}%
\bibitem [{\citenamefont {Lesinski}\ \emph {et~al.}(2009)\citenamefont
  {Lesinski}, \citenamefont {Duguet}, \citenamefont {Bennaceur},\ and\
  \citenamefont {Meyer}}]{EPJA40_121}%
  \BibitemOpen
  \bibfield  {author} {\bibinfo {author} {\bibfnamefont {T.}~\bibnamefont
  {Lesinski}}, \bibinfo {author} {\bibfnamefont {T.}~\bibnamefont {Duguet}},
  \bibinfo {author} {\bibfnamefont {K.}~\bibnamefont {Bennaceur}}, \ and\
  \bibinfo {author} {\bibfnamefont {J.}~\bibnamefont {Meyer}},\ }\href
  {\doibase 10.1140/epja/i2009-10780-y} {\bibfield  {journal} {\bibinfo
  {journal} {Eur. Phys. J. A}\ }\textbf {\bibinfo {volume} {40}},\ \bibinfo
  {pages} {121} (\bibinfo {year} {2009})}\BibitemShut {NoStop}%
\bibitem [{\citenamefont {Nakada}\ and\ \citenamefont
  {Yamagami}(2011)}]{PhysRevC.83.031302}%
  \BibitemOpen
  \bibfield  {author} {\bibinfo {author} {\bibfnamefont {H.}~\bibnamefont
  {Nakada}}\ and\ \bibinfo {author} {\bibfnamefont {M.}~\bibnamefont
  {Yamagami}},\ }\href {\doibase 10.1103/PhysRevC.83.031302} {\bibfield
  {journal} {\bibinfo  {journal} {Phys. Rev. C}\ }\textbf {\bibinfo {volume}
  {83}},\ \bibinfo {pages} {031302} (\bibinfo {year} {2011})}\BibitemShut
  {NoStop}%
\bibitem [{\citenamefont {Wang}\ \emph {et~al.}(2014)\citenamefont {Wang},
  \citenamefont {Dobaczewski}, \citenamefont {Kortelainen}, \citenamefont
  {Yu},\ and\ \citenamefont {Stoitsov}}]{PhysRevC.90.014312}%
  \BibitemOpen
  \bibfield  {author} {\bibinfo {author} {\bibfnamefont {X.~B.}\ \bibnamefont
  {Wang}}, \bibinfo {author} {\bibfnamefont {J.}~\bibnamefont {Dobaczewski}},
  \bibinfo {author} {\bibfnamefont {M.}~\bibnamefont {Kortelainen}}, \bibinfo
  {author} {\bibfnamefont {L.~F.}\ \bibnamefont {Yu}}, \ and\ \bibinfo {author}
  {\bibfnamefont {M.~V.}\ \bibnamefont {Stoitsov}},\ }\href {\doibase
  10.1103/PhysRevC.90.014312} {\bibfield  {journal} {\bibinfo  {journal} {Phys.
  Rev. C}\ }\textbf {\bibinfo {volume} {90}},\ \bibinfo {pages} {014312}
  (\bibinfo {year} {2014})}\BibitemShut {NoStop}%
\bibitem [{\citenamefont {Bayman}\ \emph {et~al.}(1969)\citenamefont {Bayman},
  \citenamefont {Bes},\ and\ \citenamefont {Broglia}}]{PhysRevLett.23.1299}%
  \BibitemOpen
  \bibfield  {author} {\bibinfo {author} {\bibfnamefont {B.~F.}\ \bibnamefont
  {Bayman}}, \bibinfo {author} {\bibfnamefont {D.~R.}\ \bibnamefont {Bes}}, \
  and\ \bibinfo {author} {\bibfnamefont {R.~A.}\ \bibnamefont {Broglia}},\
  }\href {\doibase 10.1103/PhysRevLett.23.1299} {\bibfield  {journal} {\bibinfo
   {journal} {Phys. Rev. Lett.}\ }\textbf {\bibinfo {volume} {23}},\ \bibinfo
  {pages} {1299} (\bibinfo {year} {1969})}\BibitemShut {NoStop}%
\bibitem [{\citenamefont {Dussel}\ \emph {et~al.}(1970)\citenamefont {Dussel},
  \citenamefont {Maqueda},\ and\ \citenamefont {Perazzo}}]{Dussel1970469}%
  \BibitemOpen
  \bibfield  {author} {\bibinfo {author} {\bibfnamefont {G.}~\bibnamefont
  {Dussel}}, \bibinfo {author} {\bibfnamefont {E.}~\bibnamefont {Maqueda}}, \
  and\ \bibinfo {author} {\bibfnamefont {R.}~\bibnamefont {Perazzo}},\ }\href
  {\doibase http://dx.doi.org/10.1016/0375-9474(70)90786-4} {\bibfield
  {journal} {\bibinfo  {journal} {Nucl. Phys. A}\ }\textbf {\bibinfo {volume}
  {153}},\ \bibinfo {pages} {469 } (\bibinfo {year} {1970})}\BibitemShut
  {NoStop}%
\bibitem [{\citenamefont {Hinohara}()}]{nhinprep}%
  \BibitemOpen
  \bibfield  {author} {\bibinfo {author} {\bibfnamefont {N.}~\bibnamefont
  {Hinohara}} }\href@noop {} {}\bibinfo {howpublished} {{\it et al}., in
  preparation}\BibitemShut {NoStop}%
\bibitem [{\citenamefont {Bes}\ \emph {et~al.}(1970)\citenamefont {Bes},
  \citenamefont {Broglia}, \citenamefont {Perazzo},\ and\ \citenamefont
  {Kumar}}]{Bes19701}%
  \BibitemOpen
  \bibfield  {author} {\bibinfo {author} {\bibfnamefont {D.~R.}\ \bibnamefont
  {Bes}}, \bibinfo {author} {\bibfnamefont {R.~A.}\ \bibnamefont {Broglia}},
  \bibinfo {author} {\bibfnamefont {R.~P.~J.}\ \bibnamefont {Perazzo}}, \ and\
  \bibinfo {author} {\bibfnamefont {K.}~\bibnamefont {Kumar}},\ }\href
  {\doibase DOI: 10.1016/0375-9474(70)90677-9} {\bibfield  {journal} {\bibinfo
  {journal} {Nucl. Phys. A}\ }\textbf {\bibinfo {volume} {143}},\ \bibinfo
  {pages} {1 } (\bibinfo {year} {1970})}\BibitemShut {NoStop}%
\bibitem [{\citenamefont {Clark}\ \emph {et~al.}(2006)\citenamefont {Clark},
  \citenamefont {Macchiavelli}, \citenamefont {Fortunato},\ and\ \citenamefont
  {Kr\"ucken}}]{PhysRevLett.96.032501}%
  \BibitemOpen
  \bibfield  {author} {\bibinfo {author} {\bibfnamefont {R.~M.}\ \bibnamefont
  {Clark}}, \bibinfo {author} {\bibfnamefont {A.~O.}\ \bibnamefont
  {Macchiavelli}}, \bibinfo {author} {\bibfnamefont {L.}~\bibnamefont
  {Fortunato}}, \ and\ \bibinfo {author} {\bibfnamefont {R.}~\bibnamefont
  {Kr\"ucken}},\ }\href {\doibase 10.1103/PhysRevLett.96.032501} {\bibfield
  {journal} {\bibinfo  {journal} {Phys. Rev. Lett.}\ }\textbf {\bibinfo
  {volume} {96}},\ \bibinfo {pages} {032501} (\bibinfo {year}
  {2006})}\BibitemShut {NoStop}%
\bibitem [{\citenamefont {Clark}\ and\ \citenamefont
  {Macchiavelli}(2008)}]{PhysRevC.77.057301}%
  \BibitemOpen
  \bibfield  {author} {\bibinfo {author} {\bibfnamefont {R.~M.}\ \bibnamefont
  {Clark}}\ and\ \bibinfo {author} {\bibfnamefont {A.~O.}\ \bibnamefont
  {Macchiavelli}},\ }\href {\doibase 10.1103/PhysRevC.77.057301} {\bibfield
  {journal} {\bibinfo  {journal} {Phys. Rev. C}\ }\textbf {\bibinfo {volume}
  {77}},\ \bibinfo {pages} {057301} (\bibinfo {year} {2008})}\BibitemShut
  {NoStop}%
\bibitem [{\citenamefont {Perli\ifmmode~\acute{n}\else \'{n}\fi{}ska}\ \emph
  {et~al.}(2004)\citenamefont {Perli\ifmmode~\acute{n}\else \'{n}\fi{}ska},
  \citenamefont {Rohozi\ifmmode~\acute{n}\else \'{n}\fi{}ski}, \citenamefont
  {Dobaczewski},\ and\ \citenamefont {Nazarewicz}}]{PhysRevC.69.014316}%
  \BibitemOpen
  \bibfield  {author} {\bibinfo {author} {\bibfnamefont {E.}~\bibnamefont
  {Perli\ifmmode~\acute{n}\else \'{n}\fi{}ska}}, \bibinfo {author}
  {\bibfnamefont {S.~G.}\ \bibnamefont {Rohozi\ifmmode~\acute{n}\else
  \'{n}\fi{}ski}}, \bibinfo {author} {\bibfnamefont {J.}~\bibnamefont
  {Dobaczewski}}, \ and\ \bibinfo {author} {\bibfnamefont {W.}~\bibnamefont
  {Nazarewicz}},\ }\href {\doibase 10.1103/PhysRevC.69.014316} {\bibfield
  {journal} {\bibinfo  {journal} {Phys. Rev. C}\ }\textbf {\bibinfo {volume}
  {69}},\ \bibinfo {pages} {014316} (\bibinfo {year} {2004})}\BibitemShut
  {NoStop}%
\bibitem [{\citenamefont {Sato}\ \emph {et~al.}(2013)\citenamefont {Sato},
  \citenamefont {Dobaczewski}, \citenamefont {Nakatsukasa},\ and\ \citenamefont
  {Satu\l{}a}}]{PhysRevC.88.061301}%
  \BibitemOpen
  \bibfield  {author} {\bibinfo {author} {\bibfnamefont {K.}~\bibnamefont
  {Sato}}, \bibinfo {author} {\bibfnamefont {J.}~\bibnamefont {Dobaczewski}},
  \bibinfo {author} {\bibfnamefont {T.}~\bibnamefont {Nakatsukasa}}, \ and\
  \bibinfo {author} {\bibfnamefont {W.}~\bibnamefont {Satu\l{}a}},\ }\href
  {\doibase 10.1103/PhysRevC.88.061301} {\bibfield  {journal} {\bibinfo
  {journal} {Phys. Rev. C}\ }\textbf {\bibinfo {volume} {88}},\ \bibinfo
  {pages} {061301} (\bibinfo {year} {2013})}\BibitemShut {NoStop}%
\bibitem [{\citenamefont {Sheikh}\ \emph {et~al.}(2014)\citenamefont {Sheikh},
  \citenamefont {Hinohara}, \citenamefont {Dobaczewski}, \citenamefont
  {Nakatsukasa}, \citenamefont {Nazarewicz},\ and\ \citenamefont
  {Sato}}]{PhysRevC.89.054317}%
  \BibitemOpen
  \bibfield  {author} {\bibinfo {author} {\bibfnamefont {J.~A.}\ \bibnamefont
  {Sheikh}}, \bibinfo {author} {\bibfnamefont {N.}~\bibnamefont {Hinohara}},
  \bibinfo {author} {\bibfnamefont {J.}~\bibnamefont {Dobaczewski}}, \bibinfo
  {author} {\bibfnamefont {T.}~\bibnamefont {Nakatsukasa}}, \bibinfo {author}
  {\bibfnamefont {W.}~\bibnamefont {Nazarewicz}}, \ and\ \bibinfo {author}
  {\bibfnamefont {K.}~\bibnamefont {Sato}},\ }\href {\doibase
  10.1103/PhysRevC.89.054317} {\bibfield  {journal} {\bibinfo  {journal} {Phys.
  Rev. C}\ }\textbf {\bibinfo {volume} {89}},\ \bibinfo {pages} {054317}
  (\bibinfo {year} {2014})}\BibitemShut {NoStop}%
\end{thebibliography}

\end{document}